\documentclass[aps, prd, nofootinbib, superscriptaddress, fleqn, 10pt]{revtex4-2}
\usepackage[utf8]{inputenc}
\usepackage[T1]{fontenc}
\usepackage[english]{babel}
\usepackage{amsmath, amssymb, mathtools} 
\usepackage{slashed, braket, bm}
\usepackage{float, graphicx}
\usepackage{xcolor, hyperref}
\usepackage[caption=false]{subfig}
\usepackage{booktabs}
\definecolor{blue_refs}{rgb}{0., 0., 0.85}
\hypersetup{ 
colorlinks = true,
linkcolor = blue_refs,
citecolor = blue_refs,		
filecolor = blue_refs,		
urlcolor = blue_refs
}		
\newcommand{\nt}{\notag}
\renewcommand{\d}{\mathrm{d}}
\newcommand{\e}{\mathrm{e}}
\renewcommand{\L}{\mathrm{L}}
\newcommand{\T}{{\scriptscriptstyle T}}
\newcommand{\xB}{x_{\scriptscriptstyle B}}
\newcommand{\cbar}{\bar c}
\newcommand{\qT}{\bm q_\T}

\begin{document}

\title{The impact of the TMD shape function on matching the transverse momentum spectrum in \texorpdfstring{$J/\psi$}{\textit{J/ψ}} production at the EIC}

\author{Luca Maxia}
\email{maxia@lpthe.jussieu.fr}
\affiliation{Van Swinderen Institute for Particle Physics and Gravity, University of Groningen, Nijenborgh 3, 9747 AG Groningen, The Netherlands}
\affiliation{Laboratoire de Physique Théorique et Hautes Energies (LPTHE), UMR7589, Sorbonne Université et CNRS, 4 place Jussieu, 75005 Paris, France}

\author{Daniël Boer}
\email{d.boer@rug.nl}
\affiliation{Van Swinderen Institute for Particle Physics and Gravity, University of Groningen, Nijenborgh 3, 9747 AG Groningen, The Netherlands}

\author{Jelle Bor}
\affiliation{Van Swinderen Institute for Particle Physics and Gravity, University of Groningen, Nijenborgh 3, 9747 AG Groningen, The Netherlands}
\affiliation{IJCLab, CNRS, Université Paris-Saclay, 91405 Orsay, France}

\date{\today}

\begin{abstract}
The impact of the inclusion of TMD shape functions on the transverse momentum spectrum in $J/\psi$ production at the EIC is investigated by considering the matching of the TMD factorization description at low transverse momentum with the collinear factorization description at high transverse momentum by means of the inverse-error weighting method. Despite large uncertainties from scale variations and the $J/\psi$ long-distance matrix elements, predictions for the differential cross section and its $\cos(2\phi_\psi)$ modulation are obtained. We find that physical constraints are satisfied in case a process-dependent term is included for color octet production, but not in all cases when it is excluded. These numerical results support the analytic calculations in \cite{Boer:2023zit}. Future experimental data can thus test the validity of TMD factorization in $J/\psi$ production and explore the presence of nontrivial process-dependent effects in the soft-gluon resummation. This will be crucial in the extraction of gluon unpolarized and linearly polarized TMD distributions of the proton. In addition, we suggest how the sign of the latter can be determined by investigating the presence of a node in the $\cos(2\phi_\psi)$ modulation as a function of transverse momentum.  
\end{abstract}

\date{\today}
\maketitle

\section{Introduction}
Although major steps forward have been made on the transverse momentum dependent (TMD) description of the quark distributions inside protons (\textit{e.g.}, \cite{Barone:2009hw, Barone:2015ksa, Bacchetta:2017gcc, Scimemi:2017etj, Bacchetta:2019sam, Scimemi:2019cmh, Christova:2020ahe, Bacchetta:2022awv, Moos:2023yfa, Bacchetta:2024qre, Moos:2025sal, Bacchetta:2025ara} for unpolarized and Boer-Mulders distributions), for the gluons this is still largely unknown. Therefore, it is important to identify scenarios that kinematically or dynamically favor the gluon contributions over the quark ones.
Higgs and, in particular, quarkonium production are often considered the main tools to access gluon distributions. 
Hence, experiments at the LHC and at the future electron-ion collider (EIC) will open up the opportunity to deeply explore the gluon structure of the proton through precise data on quarkonium production.
In particular, within TMD factorization, already several works~\cite{Godbole:2013bca, Boer:2016fqd, Mukherjee:2016qxa, Bacchetta:2018ivt, DAlesio:2019qpk, Chakrabarti:2022rjr, Bor:2022fga, Bacchetta:2023zir, Kishore:2024bdd, Banu:2024ywv} have pointed out observables at the EIC that can give access to such gluon distributions.
Nevertheless, proper understanding of the physics of the initial state necessitates the knowledge of the final state production process.

In recent works \cite{Echevarria:2019ynx, Fleming:2019pzj}, it has been understood that the proper factorization of quarkonium production (and decay) at small transverse momentum requires the inclusion of TMD effects in the quarkonium formation mechanism as well. Therefore, a new distribution $\Delta(z, k_\T^2)$ called TMD shape function (TMD-ShF) is introduced. It depends on $z$, the energy fraction of the quarkonium, and $k_\T$, its transverse momentum relative to the heavy quark pair from which it is produced. 
At large transverse momentum, one can perform an operator product expansion of such quantities, which connects them to their collinear counterparts used in collinear factorization. In particular, if one employs the nonrelativistic QCD (NRQCD) approach \cite{Bodwin:1994jh}, we have that the TMD-ShF is related to the long-distance matrix elements (LDMEs) by the relation
\begin{equation}
    \Delta^{[n]}_{\cal Q}(z, \bm k_\T^2) = \sum_{n'} C_{n n'}(z, \bm k_\T^2) \otimes \langle {\cal O}_{\cal Q}[n'] \rangle\ ,
\label{eq: TMDShF expansion}
\end{equation}
where $C_{nn^\prime}$ are the matching coefficients with $n$ and $n'$ being two Fock states. As commonly done, we denote a particular state using spectroscopic notation, namely $n = {}^{2S+1} L_J^{(c)}$, with $S$ being its spin, $L$ its orbital angular momentum, $J$ the total angular momentum, and $c$ its color representation. The expansion in Fock states, infinite a priori, is truncated at a certain precision in $v$, the relative velocity of the heavy quark-antiquark pair; later, we will consider states up to $v^4$. 
In Ref.~\cite{Echevarria:2019ynx} the TMD-ShF has been studied within the soft-collinear effective theory (SCET) in $pp$ collisions and for the production of color-singlet (CS) quarkonia solely. Ref.~\cite{Fleming:2019pzj} analyzed in the same formalism the $\chi_c$ decay scenario also including the color-octet (CO) channel. On the other hand, Ref.~\cite{Boer:2020bbd} and subsequently \cite{DAlesio:2023qyo, Boer:2023zit} were the first works to obtain, via a matching procedure, the large transverse momentum behavior of the TMD-ShF in semi-inclusive DIS (SIDIS). As will be very relevant in this paper, Ref.~\cite{Boer:2023zit} showed that the TMD-ShF depends on other large energy scales in the process besides the quarkonium mass, suggesting that the TMD-ShF may be separated into a universal and a process-dependent component.
In contrast, in \cite{Echevarria:2024idp}, where the $J/\psi$ production in SIDIS is studied employing the SCET formalism, the authors only retrieved the universal component, without any presence of process dependences.

The TMD formalism applies exclusively at low $q_\T$,\footnote{Here $\qT$ corresponds to the photon transverse momentum measured w.r.t.\ the proton-$J/\psi$ plane. The absolute value of $\qT$ is related to that of the $J/\psi$ transverse momentum $\bm P_{\psi\perp}$ (defined w.r.t.\ the photon-proton plane) via the relation $|\qT| = |\bm P_{\psi\perp}|/z$.} whereas for higher $q_\T$ values it should be combined with the fixed-order result obtained using collinear factorization. 
This corresponds to the so-called $W + Y$ or CSS formalism \cite{Collins:1984kg,Boglione:2014oea,Collins:2016hqq,Gamberg:2017jha,Gonzalez-Hernandez:2018ipj,Gonzalez-Hernandez:2023iso}, employed in the study of light hadron (like pions) yields in SIDIS. Besides this formalism, other ways to combine or match the two factorizations have been put forward, \textit{e.g.}, the inverse-error weighting (InEW) method, a more phenomenological approach first introduced in \cite{Echevarria:2018qyi} and recently applied to $J/\psi$ production in $pp$ collisions in \cite{Saleev:2025ryh}. In the following we also consider this phenomenological approach to present our predictions for observables accessible at the EIC, which show the impact of the TMD shape function on the matching of the full transverse momentum spectrum from low to high $q_\T$.

In particular, we evaluate the process at the partonic level at lowest nonzero order, corresponding to the $\alpha \alpha_s^2$ order in the collinear regime and to $\alpha \alpha_s$ in the TMD one. For the latter, we should account for the accuracy of the resummation of large logarithms, for which we consider the next-to-leading logarithm (NLL).
We point out that more precise results within the CS model have been presented in the literature but limited to the collinear region (\textit{e.g.}, see Ref.~\cite{Flore:2020jau}). In contrast, in this work we are primarily interested in the intermediate and low transverse momentum region, where nonperturbative transverse momentum effects are very relevant, and where the CS channel is suppressed.
Thus, for consistency we will limit our description to the NRQCD approach and the precision mentioned above.

The rest of the paper is organized as follows.
In sec.~\ref{sec: theo frame} we present the theoretical framework, discussing the main ingredients of our calculations and the matching procedure employed. In sec.~\ref{sec: cross section} we then present the numerical results for the cross section, considering different choices of the TMD-ShF. In sec.~\ref{sec: cos2phi} we discuss the numerical results and TMD-ShF dependence of the $\cos(2\phi_\psi)$ asymmetry.
In sec.~\ref{sec: conclusions} we draw our conclusions.

\section{TMD-ShF and matching procedure in electroproduction}
\label{sec: theo frame}

\begin{figure}[t]
    \centering
    \includegraphics[width=.55\linewidth, keepaspectratio]{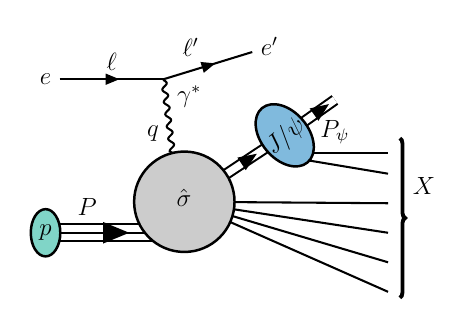}
    \caption{General representation of the SIDIS process. The gray circle denotes the hard process, the green the proton correlator, while the blue one depicts the quarkonium hadronization process (CS plus CO in this work).}
    \label{fig: SIDIS}
\end{figure}

We consider the process in Fig.~\ref{fig: SIDIS}
\begin{equation}
    \gamma^*\,(q) + p\,(P) \to J/\psi\, (P_\psi) + X\ ,
\end{equation}
where $q = \ell - \ell'$ with $q^2 = - Q^2$, and $P_\psi^2 = M_\psi^2$ with $M_\psi = 3.1~{\rm GeV}$. It is characterized by three Lorentz-invariant variables
\begin{equation}
    \xB = \frac{Q^2}{2 P \cdot q}\ , \quad y = \frac{P \cdot q}{P \cdot \ell}\ , \quad z = \frac{P \cdot P_\psi}{P \cdot q}\ ,
\end{equation}
the Bjorken-$x$, and the photon and $J/\psi$ energy fractions, respectively.
Adopting these variables, one can write the differential cross-section as~\cite{Bacchetta:2006tn, Zhang:2017dia, Bacchetta:2018ivt}
\begin{align}
    \frac{\d\sigma}{\d \xB\, \d y\, \d z\, \d^2 \bm P_{\psi\perp} } & = \frac{\alpha}{y Q^2} \frac{2}{z^2} \Big\{ \big[ 1 + (1 - y)^2 \big] F_{UUT} + 4\,(1 - y)\, F_{UUL} \nt \\ 
    & \phantom{=} + (2 - y) \sqrt{1 - y} \cos \phi_\psi\, F_{UU}^{\cos \phi_\psi} + (1 - y) \cos 2\phi_\psi\, F_{UU}^{\cos 2\phi_\psi} \Big\}\ ,
\label{eq: ep diff. cross-section}
\end{align}
where $\bm P_{\psi\perp} = P_{\psi\perp} (\cos \phi_\psi, \sin\phi_\psi)$ is the $J/\psi$ transverse momentum in a frame where the proton and photon are along the light-cone directions. Equivalently, one can identify a frame where the proton and $J/\psi$ are defined along light-cone directions and it then is the photon that acquires a transverse component $\qT$, with its modulus related to that of $\bm P_{\psi\perp}$ via the relation ${|\bm P_{\psi\perp}| = z\, |\qT|}$.
The cross section in Eq.~\eqref{eq: ep diff. cross-section} is presented in terms of structure functions, where the first two subscripts indicate the polarizations of the initial particles (electron and proton respectively), with $U$ for no polarization (unpolarized), $T$ for transverse and $L$ for longitudinal; the third subscript refers to the photon polarization; the superscript is a reminder of the presence of an azimuthal modulation. Note that for such asymmetries both photon polarizations contribute.
In this work we are interested in studying Eq.~\eqref{eq: ep diff. cross-section} at all $q_\T$, with special interest reserved to the intermediate to small-$q_\T$ region. 
Therefore, only the isotropic distribution and the $\cos2\phi_\psi$ asymmetry will be discussed here, since $F_{UU}^{\cos \phi_\psi}$ is suppressed at small $q_\T$ even upon the inclusion of the CO mechanism.

According to the CSS formalism~\cite{Collins:1984kg}, the differential cross section can be expressed as the sum of two quantities: the $Y$-term, which is constructed from contributions contained in the fixed-order (FO) result which are regular at small-$q_\T$, and the $W$-term, which incorporates the resummation of large logarithms,
\begin{equation}
    \frac{\d\sigma}{\d \xB\, \d y\, \d z\, \d \qT^2\, \d \phi_\psi} = \sum_{a, n} H^{a, [n]}(x, y, \phi_\psi; \mu) \int \frac{\d^2 \bm b_\T}{(2 \pi)^2} \e^{-i \qT \cdot \bm b_\T} W^{a, [n]}(x, z, \bm b_\T; \mu_b) + Y(x, z, \qT, \phi_\psi; \mu)\ ,
\label{eq: CSS formalism ep}
\end{equation}
where $\mu$ corresponds to the factorization scale of the process. The sum runs over different partonic and Fock states contributions. For $J/\psi$ production in SIDIS, only the gluon channel contributes at LO in $\alpha_s$ and we therefore take $a = g$ in the $W$-term henceforth. On the other hand, both quarks and gluons contribute at higher orders in $\alpha_s$ and are therefore both included in the evaluation of the regular term. In addition, we truncate the sum over $n$ to include only those states that scale as $v^4$ compared to the dominant one. Explicitly, we will hence consider only $^1 S_0^{(8)}$ and $^3 P_J^{(8)}$ contributing to Eq.~\eqref{eq: CSS formalism ep}.
The LO hard amplitudes $H^{g, [n]}$ valid at low $q_\T$ and the NLO expressions entering in the $Y$-term can be retrieved from the literature, \textit{e.g.}, in \cite{Boer:2021ehu,Boer:2023zit} for the former and \cite{Kniehl:2001tk, Sun:2017nly} for the latter.
Moreover, note that in Eq.~\eqref{eq: CSS formalism ep} the variable $x$, which corresponds to the exact light-cone fraction at small-$q_\T$, has been introduced. It is directly proportional to $\xB$ via
\begin{equation}
    x = \xB/ \hat x_{\rm max}(\qT)\ ,
\label{eq: x-xB relation}
\end{equation}
where $\hat x_{\rm max}$ is the maximal value of the partonic Bjorken-$\hat x$
\begin{equation}
    \hat x_{\rm max}(\qT) = \frac{Q^2}{M_\psi^2 + Q^2 + 2 M_\psi |\qT|}\ ,
\label{eq: xmax function}
\end{equation}
Note that in Refs.~\cite{Boer:2020bbd, Boer:2023zit} $\hat x_{\rm max}$ was presented as independent of $q_\T$ since the authors' interest was primarily on the kinematical region where $q_\T \ll M_\psi, Q$. However, Eq.~\eqref{eq: xmax function} is needed to properly evaluate the cross section at all $q_\T$.

The $W$-term in Eq.~\eqref{eq: CSS formalism ep} is already presented in $\bm b_\T$ space, the Fourier conjugate of $\bm q_T$. 
The explicit form of the integrand $W^{g, [n]}$ for $b_\T^2 \ll \Lambda_{\rm QCD}^2$ is
\begin{equation}
     W^{g, [n]}(x, z, \bm b_\T; \mu_b, \mu) = \sum_{a'} C_{g a'} \otimes f_1^{a'}(x; \mu_b)\, \sum_{n'} C_{n n'} \otimes \langle {\cal O}_{\cal \psi}[n'] \rangle(z; \mu_b)\, \e^{-S_{\rm pert}(\mu_b, \mu)}\ ,
\label{eq: W-term}
\end{equation}
where we have introduced the scale $\mu_b = b_0/|\bm b_\T|$, with ${b_0 = 2\e^{-\gamma_E} \approx 1.123}$. 
A few comments on Eq.~\eqref{eq: W-term}:
\begin{enumerate}
    \item While the PDFs can be evaluated at $\mu_b$, the LDMEs are extracted at a specific scale, usually $\mu_\psi \sim M_\psi$. However, the evolution of the LDMEs is off diagonal~\cite{Butenschoen:2019lef, Echevarria:2024idp}. More specifically for the case under study, the off-diagonal terms relevant for $^1 S_0$ and $^3 P_J$ are respectively related to $^1 P_1$ states and $D$ waves, which are negligible for $J/\psi$. Thus, in the evolution from $\mu$ to $\mu_b$, the magnitude of the distribution does not evolve, and we can approximate ${\langle {\cal O}_{\cal \psi}[n'] \rangle (\mu_b) \approx \langle {\cal O}_{\cal \psi}[n'] \rangle (\mu_\psi)}$ in Eq.~\eqref{eq: W-term}. On the other hand, the PDFs at scale $\mu_b$ are evaluated using LHAPDF grids~\cite{Buckley:2014ana}.
    \item The perturbative Sudakov factor, which resums large logarithms in $\bm b_\T$, receives contributions from the (incoming) gluon and (outgoing) $c\cbar$ pair, and it can be parameterized in the general form
    \begin{equation}
        S_{\rm pert} = \int_{\mu_b^2}^{\mu^2} \frac{\d \eta^2}{\eta^2} \left[ A_g\big(\alpha_s(\eta)\big)\, \log \frac{\mu^2}{\eta^2} + B_g\big(\alpha_s(\eta)\big) \right] + \int_{\mu_b^2}^{\mu^2} \frac{\d \eta^2}{\eta^2} B_{\rm CO} \big(\alpha_s(\eta)\big)\ ,
    \label{eq: Sudakov definition}
    \end{equation}
    with the coefficients $A$ and $B$ that can be expanded in series of $\alpha_s$ according to
    \begin{equation}
    \begin{aligned}
        A\big(\alpha_s(\eta)\big) & = \sum_{k=1}^\infty \left( \frac{\alpha_s(\eta)}{\pi} \right)^k A^{(k)} \ , \\
        B\big(\alpha_s(\eta)\big) & = \sum_{k=1}^\infty \left( \frac{\alpha_s(\eta)}{\pi} \right)^k B^{(k)} \ .
    \end{aligned}
    \end{equation}
    The gluon contributes to both single- and double-logarithms;
    at NLL accuracy \cite{Catani:1988vd}, $A_g$ at $2$-loops is
    \begin{equation}
        A_g^{(1)} = \frac{C_A}{2}\ , \quad A_g^{(2)} = \frac{C_A}{4} \left[ C_A \left( \frac{67}{18} - \frac{\pi^2}{6} \right) - \frac{5}{9}n_f \right]\ ,
    \end{equation}
     and $B_g$ at $1$-loop
    \begin{equation}
        B_g^{(1)} = - \frac{C_A}{2} \frac{\beta_0}{6}\ ,
    \end{equation}
    where $\beta_0 = 11 - 2n_f/3 = 23/3$ for $n_f = 5$.
    On the other hand, the $J/\psi$ only contributes to the single logarithm~\cite{Sun:2012vc, Echevarria:2019ynx, Fleming:2019pzj, Boer:2023zit}, with its contribution at $1$-loop given by
    \begin{equation}
        B_\psi^{(1)} = - \frac{C_A}{2}\ .
    \label{eq: Bpsi definition}
    \end{equation}
    In addition, as discussed in \cite{Boer:2023zit}, another soft contribution induced by the process may be present, given by
    \begin{equation}
         B_{ep}^{(1)} = \frac{C_A}{2}\log\frac{M_\psi^2 + Q^2}{M_\psi^2}\ .
    \label{eq: Bep definition}
    \end{equation}
    An analogous term was also found in open heavy-quark production \cite{Zhu:2013yxa}, with Eqs.~\eqref{eq: Bpsi definition} and~\eqref{eq: Bep definition} corresponding to the bound-state limit of the pair.
    Note that compared to \cite{Boer:2023zit} we have dropped the dependence on $\mu$ of $B_{ep}$. Since $B_\psi$ does not evolve with $\mu$, neither should $B_{ep}$, and the overall contribution to the single logarithms driven by the presence of a CO state in the final state is then simply identified by ${B_{\rm CO} = B_\psi + B_{ep}}$; see appendix~\ref{app: anom dim} for more details. 
    \item The two matching (or Wilson) coefficients $C_{a a'}$ and $C_{n n'}$ can also be organized in powers or $\alpha_s$
    \begin{equation}
    \begin{aligned}
        C_{a a'} (x; \alpha_s(\mu_b)) & = \sum_{k=1}^\infty \left( \frac{\alpha_s(\mu_b)}{\pi} \right)^k C_{a a'}^{(k)} (x) \ , \\
        C_{n n'} (z; \alpha_s(\mu_b)) & = \sum_{k=1}^\infty \left( \frac{\alpha_s(\mu_b)}{\pi} \right)^k C_{a a'}^{(k)} (z) \ ;
    \end{aligned}
    \end{equation}
    only their lowest order is required up to NLL, for which the coefficients reduce to delta functions
    \begin{equation}
        C_{aa'}^{(0)}(x) = \delta_{aa'}\, \delta(1 - x)\ , \quad C_{nn'}^{(0)}(z) = \delta_{nn'}\, \delta(1 - z)\ .
    \label{eq: coefficients}
    \end{equation} 
    Note that for $F_{UU}^{\cos 2\phi_\psi}$ the first nontrivial coefficients, denoted by $G_{aa'}$ to distinguish them from the $C_{aa'}$ of the isotropic component, are found at $\alpha_s$ \cite{Catani:2010pd}
    \begin{equation}
        G_{aa'}^{(0)}(x) = 0\ , \quad 
        G_{ag}^{(1)}(x) = -C_A\, \left(1 - \frac{1}{x} \right)\ , \quad G_{aq}^{(1)}(x) = -C_F\, \left(1 - \frac{1}{x} \right)\ .
    \label{eq: coefficients - h1Tperp}
    \end{equation}
    \item The symbol $\otimes$ denotes a convolution in either the parton light-cone fraction $x$ or the $c\cbar$ pair energy fraction $z$
    \begin{equation}
    \begin{aligned}
        C_{a a'} \otimes f^{a'} (x) & = \int_x^1 \frac{\d x'}{x'}\, C_{a a'}(x')\, f^{a'}(x/x')\ , \\
        C_{n n'} \otimes \langle {\cal O}_{\cal \psi}[n'] \rangle (z) & = \int_z^1 \frac{\d z'}{z'}\, C_{n n'}(z')\, \langle {\cal O}_{\cal \psi}[n'] \rangle\, \delta(1 - z/z')\ . \\
    \end{aligned}
    \end{equation}
    \item Eq.~\eqref{eq: W-term} depends on three scales which can be varied to estimate higher-order corrections. More specifically, we substitute $\mu_b \to C_1 \mu_b$ and $\mu \to C_2 \mu$ in the Sudakov factor, and $\mu_b \to C_3 \mu_b$ in the TMD-PDF \cite{Collins:1984kg,Melis:2015ycg}. 
    We consider variations by a factor $2$, \textit{i.e.}~$C_i \in [1/2, 2]$, with the central values corresponding to $C_i = 1$. 
    All these variations also affect the quantities presented above, which should be changed by adding the following terms
    \begin{equation}
        \Delta A_g^{(1)} = 0 \ , \quad \Delta A_g^{(2)} = \frac{C_A}{4}\, \beta_0 \log C_1\ , \quad \Delta B_g^{(1)} = - C_A \log \left( \frac{C_2}{C_1} \right)\ .
    \end{equation}
    Note that a similar variation should also apply to the scale at which the LDMEs are extracted.
    However, it seems reasonable to assume that the errors given in each LDME set already account for, to some extent, this scale uncertainty. Hence, in the following we will consider only the parameter errors given in the corresponding LDME set.
\end{enumerate}

\noindent
By taking $A$ and $B$ expressions and the $\alpha_s$ expansion at $1$-loop, one obtains the Sudakov factor evaluated at $1$-loop accuracy
\begin{equation}
    S_{\rm pert}^{1\text{-}{\rm loop}}(\mu^2, \mu_b^2) = - \frac{4}{\beta_0} A_g^{(1)} \left[ \log \left(\frac{\mu^2}{\mu_b^2}\right) - \left( \frac{B_g^{(1)} + B_\psi^{(1)} + B_{ep}^{(1)}}{A_g^{(1)}} + \L_\mu \right)\log\left(\frac{\L_\mu}{\L_{\mu_b}}\right) \right]\ ,
\label{eq: Sudakov at 1-loop}
\end{equation}
whereas at NLL accuracy we have
\begin{align}
    S_{\rm pert}^{\rm NLL}(\mu^2, \mu_b^2) = S_{\rm pert}^{1\text{-}{\rm loop}} - \frac{16}{\beta_0^2} A_g^{(2)} \left[ \log \left(\frac{\L_\mu}{\L_{\mu_b}} \right) - \frac{1}{\L_{\mu_b}} \log\left(\frac{\mu^2}{\mu_b^2}\right) \right]\ , 
\label{eq: Sudakov at NLL}
\end{align}
where $\L_\eta = \log\frac{\eta^2}{\Lambda_{\rm QCD}^2}$. 
In the next sections we will evaluate our findings using Eq.~\eqref{eq: Sudakov at NLL}, but we have checked that the $1$-loop result accounts for the majority of the NLL one.

Independently from the accuracy considered, to ensure that the perturbative expression in Eq.~\eqref{eq: W-term} is not used outside its range of validity, we need to
exclude the large-$b_\T$ region from the integral. In addition, to prevent the scale $\mu_b$ from becoming larger than the hard scale, causing artificial Sudakov enhancement, the small-$b_\T$ region should also be excluded from the integral.
We therefore introduce $b_{\rm max} = 1.5~{\rm GeV}^{-1}$ and $b_{\rm min} = b_0/\mu$, two parameters used in the replacement ${\mu_b \to \mu_{b_*}^\prime}$. 
This substitution is not unique, and different prescriptions that perform this operation can be found in the literature. Here we follow~\cite{Collins:1984kg, Collins:2016hqq, Bor:2022fga}\footnote{Note that in the notation of ~\cite{Collins:2016hqq} we are using $b_c(b_*(b_T))$ (with $C_5=1$) rather than their $b_*(b_c(b_T))$ in order to ensure that $S_{\rm pert}(0)=1$ and the value of $b_{\min} \equiv b_c(b_*(0))$ is then independent of $b_{\max}$, unlike $b_*(b_c(0))$.} for which
\begin{equation}
    b_* = \frac{b_\T}{\sqrt{1 + (b_\T/b_{\rm max} \big)^2}}
\end{equation}
and
\begin{equation}
    \mu_{b_*}^\prime = \frac{b_0}{\sqrt{b_*^2 + b_{\rm min}^2}}\ .
\end{equation}
Upon replacement of $\mu_b \to \mu_{b_*}^\prime$, a nonperturbative quantity must be introduced to restore the correct $b_\T$ dependence above $b_{\rm max}$:
\begin{equation}
    W^{g, [n]}(x, z, \bm b_\T; \mu_b) = W^{g, [n]}(x, z, \bm b_*; \mu_{b_*}^\prime) \, \e^{-S_{\rm NP}}\ ,
\end{equation}
where $W^{g, [n]}(x, z, \bm b_*; \mu_{b_*}^\prime)$ is given by the perturbative expression in Eq.~\eqref{eq: W-term}.
 Following in part \cite{Bor:2022fga, Aybat:2011zv}, we parameterize the nonperturbative Sudakov factor according to 
\begin{equation}
    S_{\rm NP} = \left( A_{\rm NP} \log\frac{\mu}{\mu_{\rm NP}} + B_{\rm NP} + g_\psi \right) b_T^2\ ,
\label{eq: nonperturbative Sudakov}
\end{equation}
where $\mu_{\rm NP} = 1.6~{\rm GeV}$, $A_{\rm NP} \in [0.05, 0.8]~{\rm GeV}^2$ with its central value located at $0.414~{\rm GeV}^2$, and $B_{\rm NP} = \frac{9}{8} \Big[ g_1 \big(1 + 2 g_3 \log\frac{10\, x\, x_c}{x_c + x}\big) - g_2 \log(2) \Big]$, with ${g_1 = 0.201~{\rm GeV}^2}$, ${g_2 = 0.184~{\rm GeV}^2}$, ${g_3 = -0.129~{\rm GeV}^2}$, and ${x_c = 0.009}$. Here the range of $A_{\rm NP}$ applies for one gluon TMD, as explained in \cite{Scarpa:2019fol,Bor:2025ztq}, whereas there is no such factor included for the TMD-ShF, in line with the absence of a double logarithmic term in $S_{\rm pert}$ for it. In addition, the parameterization of \cite{Aybat:2011zv} that is used to obtain $B_{\rm NP}$ refers to the case for two quark TMDs. This is the origin of the additional factor $C_A/(2C_F)=9/8$ in $B_{\rm NP}$, which differs by a factor $1/2$ from the one used in \cite{Bor:2022fga}. We include a separate factor $g_\psi$ for the TMD-ShF. Upon equating $g_\psi$ and the ($x$-dependent) $B_{\rm NP}$ used here, one recovers the $B_{\rm NP}$ of \cite{Bor:2022fga}. However, since we do not have any information about $g_\psi$, we make the assumption that it is a constant since the TMD region of the process is dominated by $z = 1$, suppressing any possible $z$ dependence in $S_{\rm NP}$.\footnote{Another point worth stressing is the possible $Q$ dependence of $g_\psi$, in other words, the possibility that it depends not only on the quarkonium involved, but also on the process considered. 
The same question was already encountered and pointed out for open-quark production, \textit{e.g.}~see \cite{Catani:2014qha}. 
Hence, the presence of nontrivial process dependence effects in the nonperturbative Sudakov factor cannot be ruled out a priori, as they might survive in the bound state limit of the quark pair.
However, in the following we proceed under the assumption that the process factorizes (leading to Eq.~\eqref{eq: W-term}), with any process-dependent  contributions to $S_{\rm NP}$ arising from $g_\psi$ being suppressed. While this assumption may be false, we nonetheless expect uncertainties on $g_\psi$ to be less relevant than the other ones considered in this paper.}

To apply Eq.~\eqref{eq: CSS formalism ep} reliably one should precisely know the asymptotic behavior of the FO cross section. 
However, our knowledge of its $z$ dependence is currently restricted to the $z \to 1$ limit, which is the dominant contribution for $J/\psi$ production in SIDIS. Therefore, in this work we evaluate the matching with the InEW method \cite{Echevarria:2018qyi}, allowing us to bypass this issue. Below we discuss some of the key ingredients of this approach, but for a more exhaustive discussion we refer the reader to \cite{Echevarria:2018qyi}.
Depending on the kinematical region considered, we can approximate, order by order in $\alpha_s$, the differential cross section in Eq.~\eqref{eq: CSS formalism ep} with either the $W$-term or the FO calculation:
\begin{equation}
    \frac{\d\sigma}{\d \xB\, \d y\, \d z\, \d \qT^2\, \d \phi_\psi}\bigg\vert_{q_\T \ll \mu(M_\psi, Q)} = \sum_{n} H^{g, [n]}(\mu) \int \frac{\d^2 \bm b_\T}{(2 \pi)^2} \e^{-i \qT \cdot \bm b_\T} W^{g, [n]}(\bm b_\T; \mu_b) + {\cal O} \left( \frac{\qT^2}{\mu^2} \right) + {\cal O} \left( \frac{m^2}{\mu^2} \right)\ ,
\label{eq: cross-section TMD approx}
\end{equation}
and
\begin{equation}
    \frac{\d\sigma}{\d \xB\, \d y\, \d z\, \d \qT^2\, \d \phi_\psi}\bigg\vert_{q_\T \gg \Lambda_{\rm QCD}} = \frac{\d\sigma}{\d \xB\, \d y\, \d z\, \d \qT^2\, \d \phi_\psi}\bigg\vert_{\rm FO} + {\cal O} \left( \frac{m^2}{\qT^2} \right) \ ,
\label{eq: cross-section collinear approx}
\end{equation}
where the energy $m$ corresponds to the scale below which the perturbative expansion in the collinear framework becomes unreliable. 
According to the InEW method, the exact differential cross section can be approximated according to
\begin{align}
    \frac{\d\sigma}{\d \xB\, \d y\, \d z\, \d \qT^2\, \d \phi_\psi} & \approx \omega_1\,  \sum_{n} H^{g, [n]} \int \frac{\d^2 \bm b_\T}{(2 \pi)^2} \e^{-i \qT \cdot \bm b_\T} W^{g, [n]} + \omega_2\, \frac{\d\sigma}{\d \xB\, \d y\, \d z\, \d \qT^2\, \d \phi_\psi}\bigg\vert_{\rm FO}\ ,
\label{eq: cross section with InEW approx}
\end{align}
where $\omega_1$ and $\omega_2$ are two weights that fulfill the condition $\omega_1 + \omega_2 = 1$. Although, strictly speaking, Eq.~\eqref{eq: cross section with InEW approx} includes the TMD expression also for large $q_T$, \textit{i.e.}\ outside its range of validity and where it even can become negative, the weights will be such that this will not impact the results.
The weights can be constructed from the errors in Eqs.~\eqref{eq: cross-section TMD approx} and~\eqref{eq: cross-section collinear approx}, namely
\begin{equation}
\begin{aligned}
    \omega_1 & = \frac{\Delta W^{-n}}{\Delta W^{-n} + \Delta {\rm FO}^{-n}}\ , \\
    \omega_2 & = \frac{\Delta {\rm FO}^{-n}}{\Delta W^{-n} + \Delta {\rm FO}^{-n}} \ ,
\end{aligned}
\label{eq: weights InEW}
\end{equation}
with
\begin{equation}
\begin{aligned}
    \Delta W & = \left( \frac{\qT^2}{\mu^2} + \frac{m^2}{\mu^2} \right)\, \frac{\d\sigma}{\d \xB\, \d y\, \d z\, \d \qT^2\, \d \phi_\psi} \ ,\\
    \Delta {\rm FO} & = \frac{m^2}{\qT^2} \left( 1 + \log\frac{\sqrt{M_\psi^2 + Q^2}}{M_\psi} + \log\frac{\sqrt{\mu^2 + \qT^2}}{\qT}\right)\, \frac{\d\sigma}{\d \xB\, \d y\, \d z\, \d \qT^2\, \d \phi_\psi}\ .
\end{aligned}
\label{eq: scheme theoretical errors}
\end{equation}
Note that, while
the cross sections on the right-hand sides correspond to the (unknown) exact cross section, the weights in Eq.~\eqref{eq: weights InEW} are independent of it. 
The last term in $\Delta {\rm FO}$ is needed to improve the reliability of the method for all $\qT$ values, and especially when $q_\T \ll \mu$ since it corresponds to the typical leading logarithm of the FO cross section. 
However, in contrast to Ref.~\cite{Echevarria:2018qyi} and inspired by the divergent behavior found in Ref.~\cite{Boer:2023zit}, we modified this term by including a logarithm of hard scales, which becomes relevant for high-$Q$ values, and removed the square from the last term ($\log\frac{\sqrt{\mu^2 + \qT^2}}{\qT}$). 
Finally, there are two choices left in the definition of the weights. The first is the power $n$, which dictates how fast the transition from $W$ to the FO (and vice versa) happens. The second is the exact value of $m$, usually set approximately at $1~{\rm GeV}$ ($\approx$ the proton mass), but it is more generally an unknown function of $\Lambda_{\rm QCD}$. 
Note that the values of $n$ and $m$ cannot be completely arbitrary, as, for instance, they must be chosen to ensure that the TMD contribution becomes negligible for $q_\T \gtrsim \mu$, which guarantees that the cross section is fully described by the FO curve at high $q_\T$.
To fulfill this requirement at all $Q$ values, and 
in line with \cite{Echevarria:2018qyi}, we take $n=2$ and $m = 1~{\rm GeV}$.

\begin{figure}[t]
  \centering    
  \includegraphics[width=.6\linewidth, keepaspectratio]{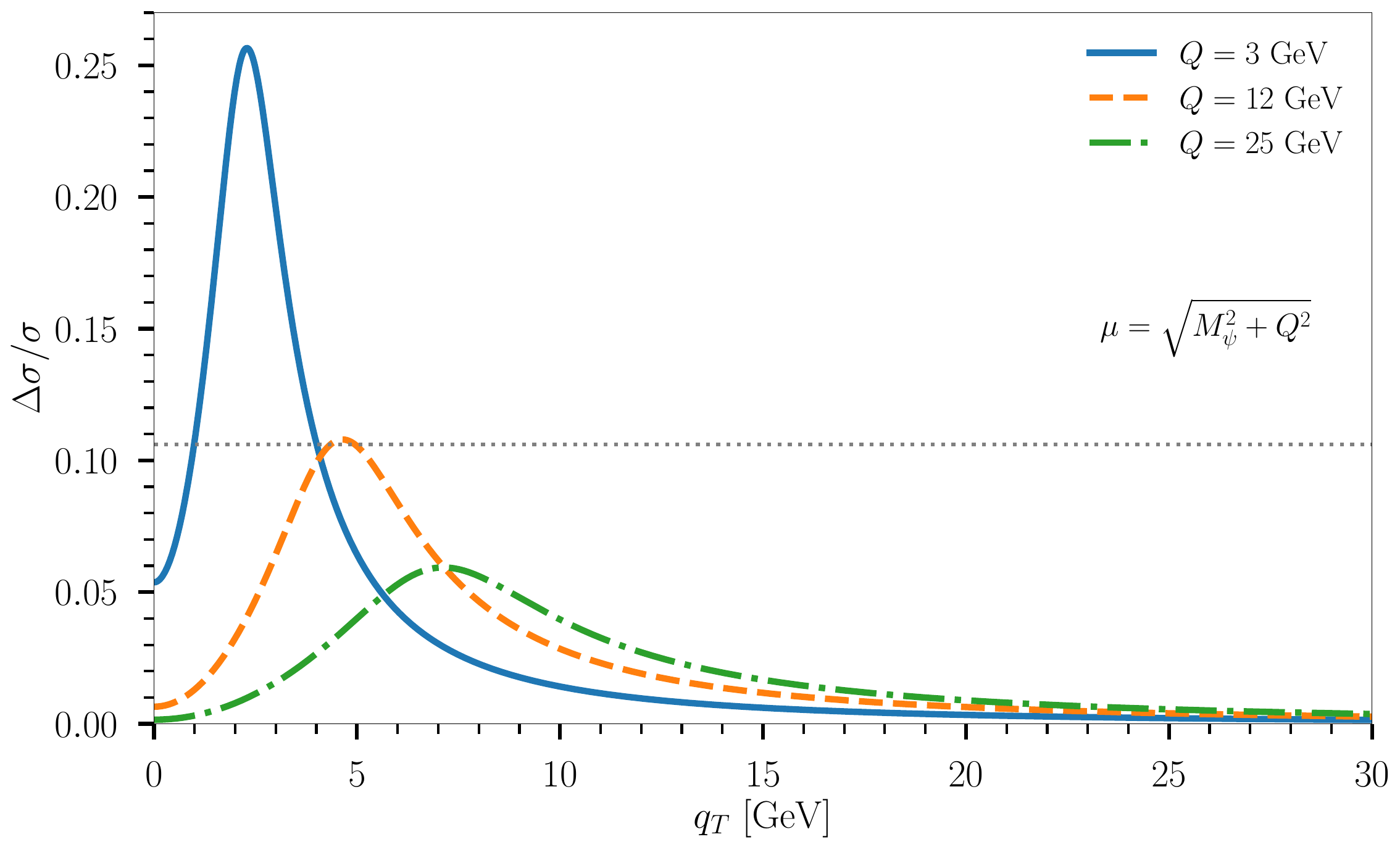}
  \caption{\it Dependence of the matching scheme error $\Delta \sigma$ on $q_\T$ for different choices of $Q$. We have considered $n=2$ and $m = 1~{\rm GeV}$, and taken $\mu = \sqrt{M_\psi^2 + Q^2}$ as the hard scale. The horizontal dotted line corresponds to the chosen threshold of $\Delta\sigma/\sigma = 10.6\%$.}
  \label{fig: InEW error}
\end{figure}

The approximation in Eq.~\eqref{eq: cross section with InEW approx} also introduces an error $\Delta \sigma$
\begin{equation}
    \Delta \sigma = \frac{\Delta W\, \Delta {\rm FO}}{\sqrt{\Delta W^2 + \Delta {\rm FO}^2}} \ ,
\label{eq: intrinsic error InEW}
\end{equation}
which is obtained by the propagation of independent errors. To estimate this error, we evaluate $\Delta W$ and $\Delta {\rm FO}$ by using the approximated cross section given in Eq.~\eqref{eq: cross section with InEW approx}, since the difference would be of higher order.
In the coming sections, we will combine this error with those obtained from other sources, including scale variations and LDME uncertainties.
The error in Eq.~\eqref{eq: intrinsic error InEW} identifies two scenarios displayed in Fig.~\ref{fig: InEW error}. 
For a sufficiently high $Q$ we have a smooth transition from the collinear to TMD factorizations (green dash-dotted line), which corresponds to a $q_\T$ region where the associated errors of both frameworks are still relatively small. 
On the other hand, at lower values of $Q$ we might force the transition in a $q_\T$ region where neither framework is trustworthy (solid blue line). The transition from the former case to the latter is decided by a certain threshold, directly related to another threshold $\Delta T$ on the relative errors of the individual frameworks. More specifically, the threshold for the matching error will be $\Delta \sigma/\sigma < \Delta T/\sqrt 2$, which corresponds to the case where frameworks exceed $\Delta T$ (note that the $\sqrt{2}$ arises from the observation that for $\Delta W = \Delta {\rm FO}$: $\Delta \sigma = \Delta W/\sqrt 2$). In Fig.~\ref{fig: InEW error} we have chosen $\Delta T = 15\%$, and therefore $\Delta \sigma/\sigma \lesssim 10.6\%$. In the next section, adopting the same threshold, we will discuss the matching in both cases.

\section{Isotropic \texorpdfstring{$J/\psi$}{\textit{J/ψ}} production in unpolarized collisions}
\label{sec: cross section}

We first investigate the isotropic contribution to Eq.~\eqref{eq: ep diff. cross-section} at fixed $Q$, $\xB$ and integrated over $z$, namely
\begin{equation}
    \int_{\rm z_{\rm min}}^1 \d z\frac{\d \sigma}{\d \xB\, \d Q\, \d q_\T\, \d z}\bigg|_{\xB = \overline{x}_{\scriptscriptstyle B}, Q = \overline Q} = \frac{4 \pi \alpha\, q_\T}{\overline Q^3} \int_{\rm z_{\rm min}}^1 \d z \Big\{ \big[ 1 + (1 - y)^2 \big]\, F_{UUT} + 4(1 - y)\, F_{UUL} \Big\} \ .
\label{eq: differential cross section - fixed}
\end{equation}
Compared to the production of light hadrons in SIDIS the choice of $z$ is now more restricted. At low transverse momentum the produced $c \bar c$ pair takes, by momentum conservation, all the energy of the photon-parton system. Consequently, its energy fraction is $z_{c\bar c} = 1$ and, due to the approximation $M_{c\cbar} \approx M_\psi$, we also have $z \approx z_{c\bar c} = 1$. Hence, a region $z_{\rm min} < z < 1$ will always be dominated by $z \approx 1$ at sufficiently low $q_\T$ independently from the specific choice of $z_{\rm min}$; in particular, in the following we will take $z_{\rm min} = 0$, considering the most extreme case where we integrate over the whole $z$ region.\footnote{As a side note, we point out that if one takes $z_{\rm min} > 0$ will primarily affect the fall-off of the curve at high-$q_\T$, dictated by the FO prediction.} 
Of course, the approximation is valid up to corrections of order $q_\T^2/M_\psi^2$, which are expected in the complete $z$ dependence of the TMD-ShF, as also suggested by~\cite{Beneke:1997qw, Beneke:1999gq}.
Such contributions must thus be included if one considers $z_{\rm min} < z < z_{\rm max}$ with $z_{\rm max} < 1$. However, in this work we discuss only the $z_{\rm max} = 1$ case and leave studies on the $z$-dependence of the TMD-ShF to the future.
Another important difference with \cite{Boglione:2014oea, Gonzalez-Hernandez:2023iso} resides in Eq.~\eqref{eq: x-xB relation}: the presence of a massive final state not only causes $x$ to differ from $\xB$ but makes this difference $q_\T$ dependent.
Therefore, each bin of $\xB$ (or equivalently $W_{\gamma p}$ - the energy of the photon-proton system) will probe a light-cone fraction $x \gtrsim \xB$, and their difference increases linearly with $q_\T$, with a slope determined by $Q$.

\begin{figure}[t]
  \centering    
  \subfloat[\label{subfig:a-match_140}]{\includegraphics[width=.47\linewidth, keepaspectratio]
  {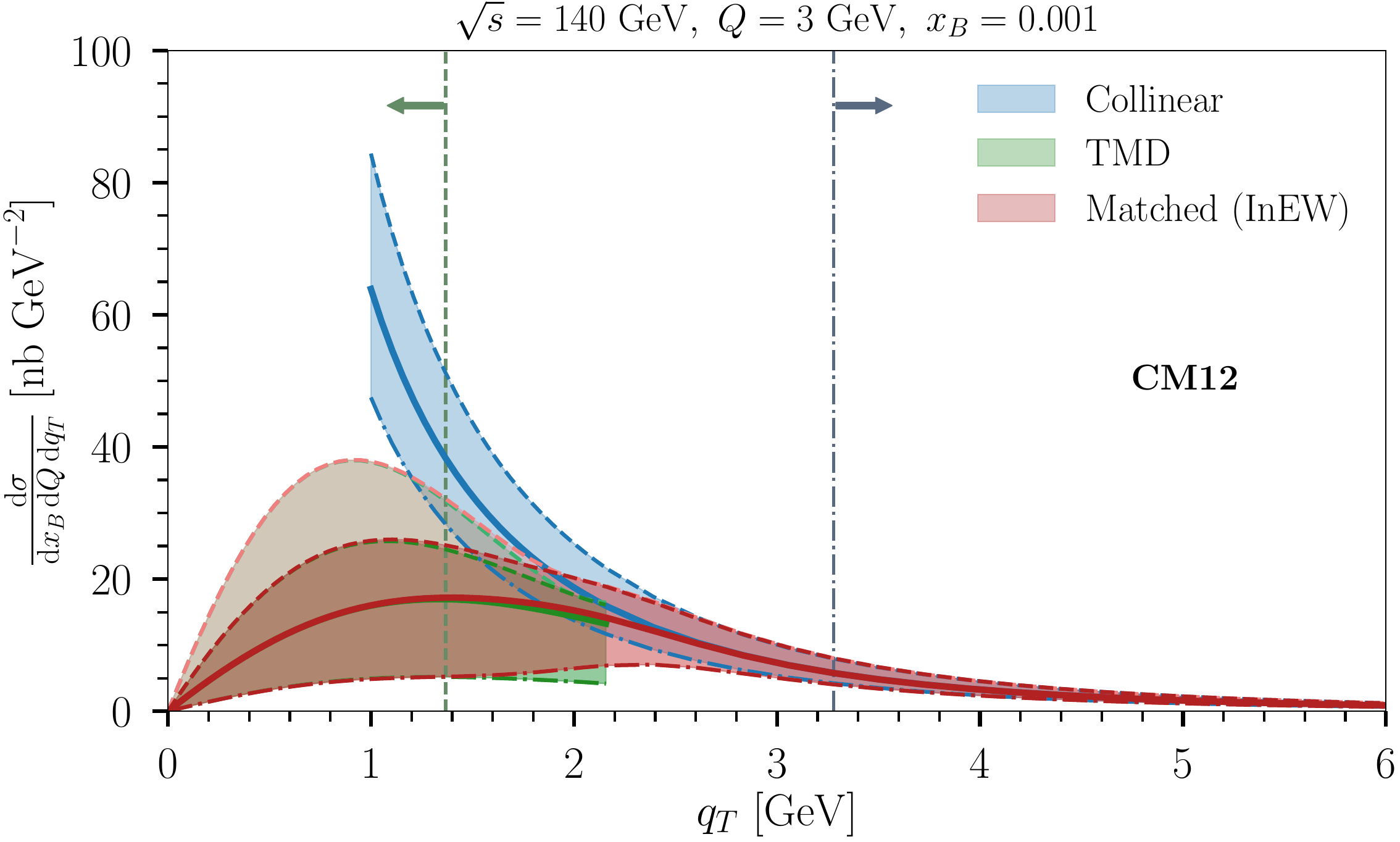}}
  \hfill
  \subfloat[\label{subfig:b-match_140}]
  {\includegraphics[width=.47\linewidth, keepaspectratio]{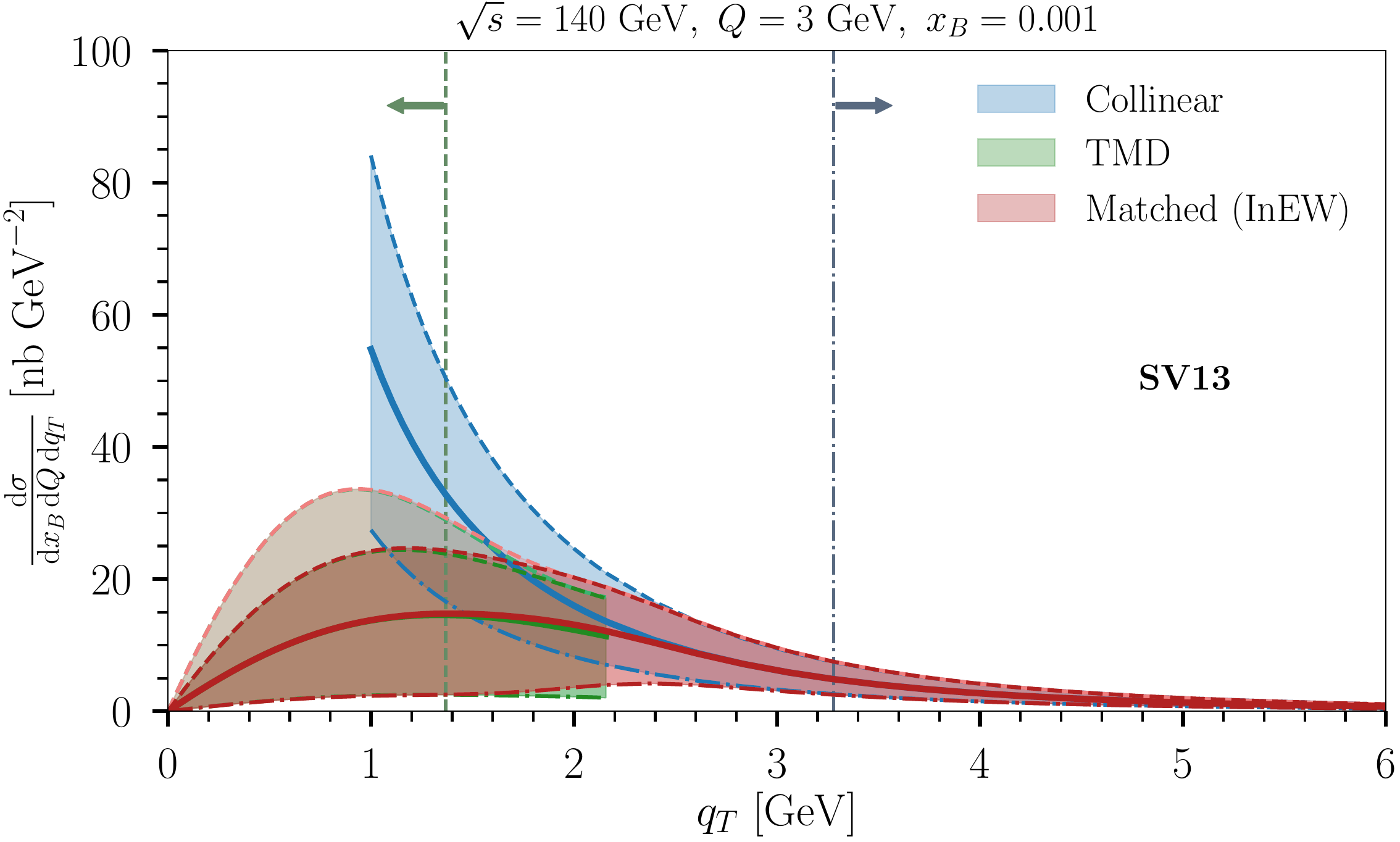}}\\
  \subfloat[\label{subfig:c-match_140}]{\includegraphics[width=.47\linewidth, keepaspectratio]{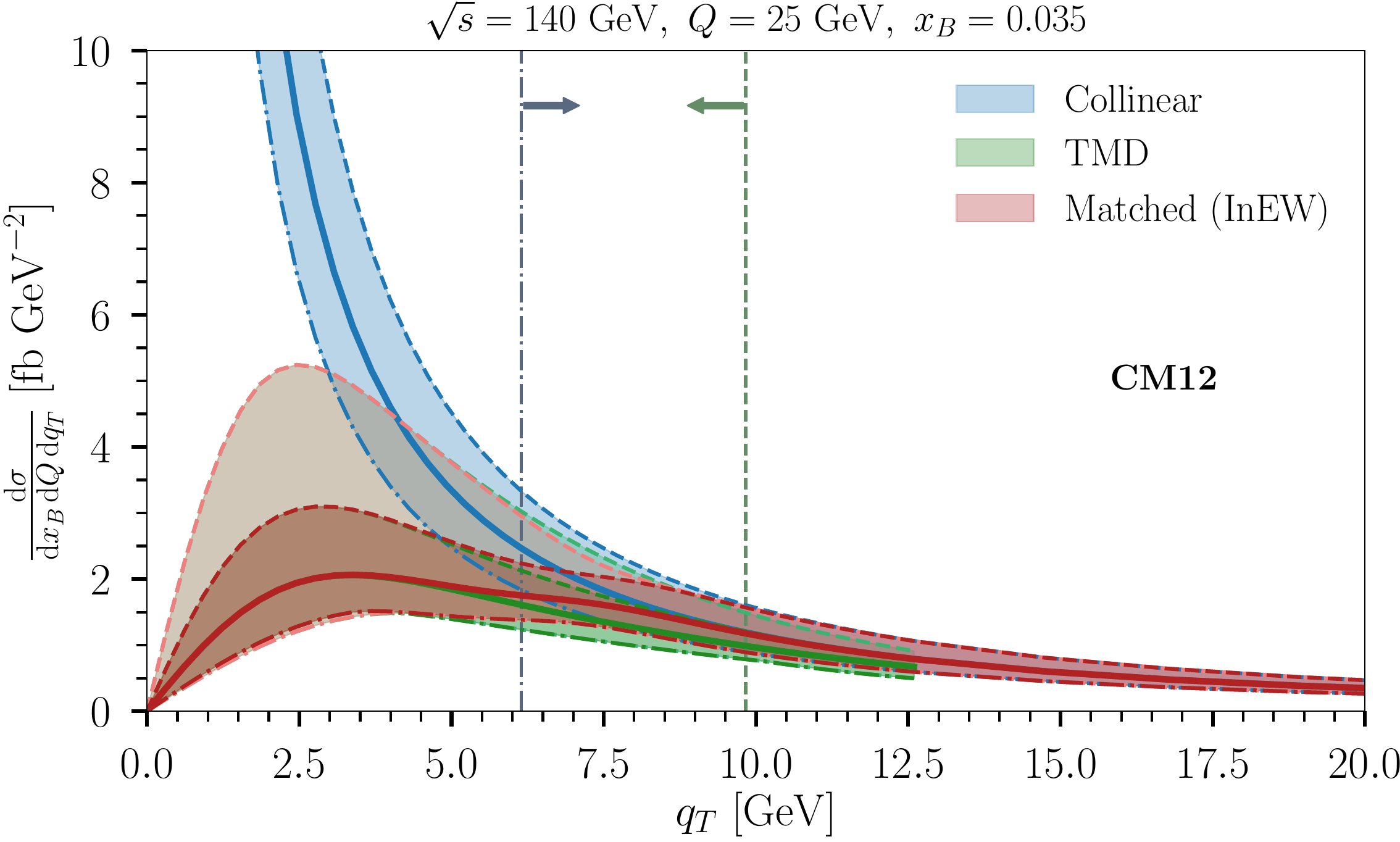}}
  \hfill
  \subfloat[\label{subfig:d-match_140}]
  {\includegraphics[width=.47\linewidth, keepaspectratio]{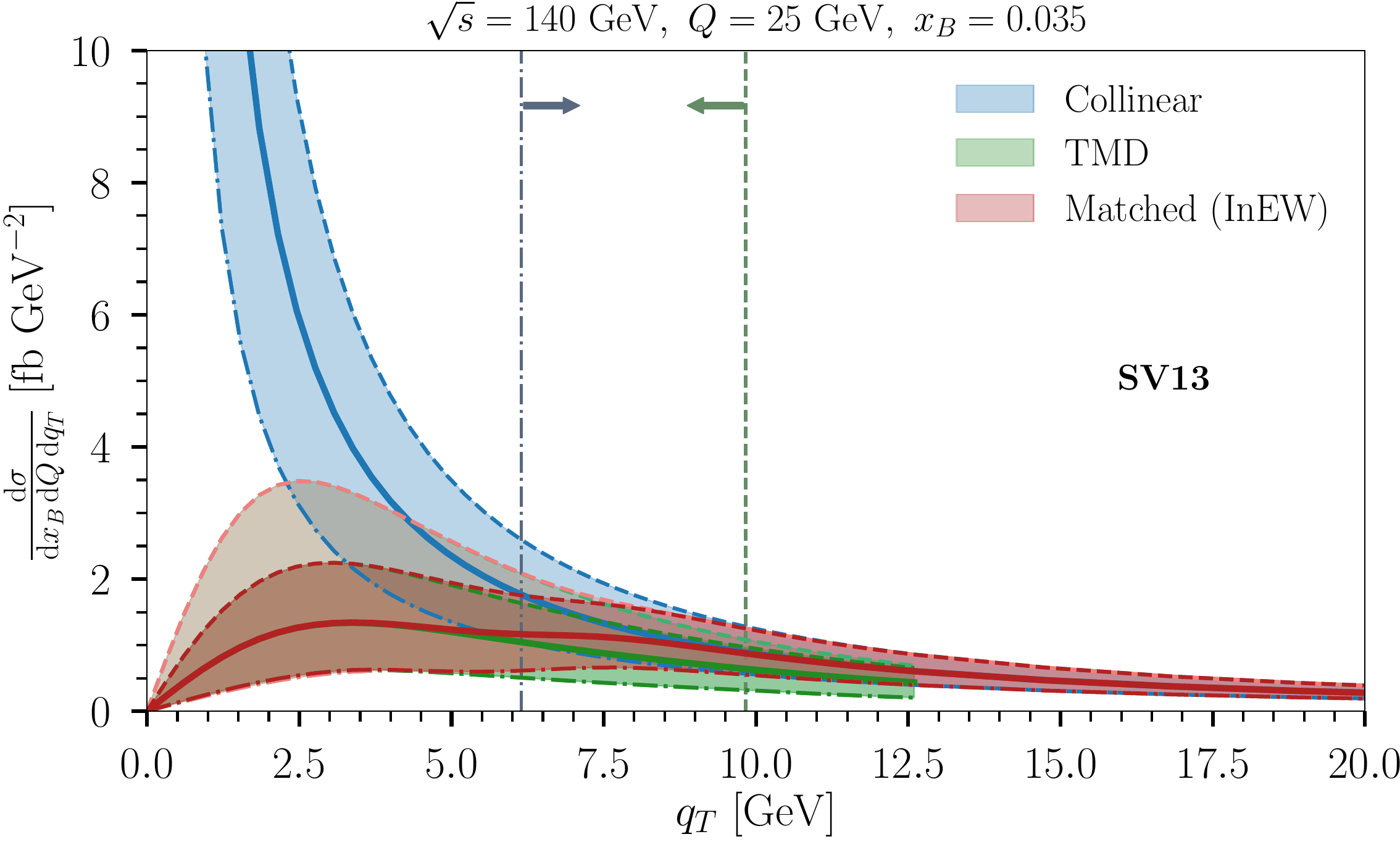}}
  \caption{\it Differential cross section with respect to $q_\T$ for $\sqrt{s} = 140~{\rm GeV}$. Figs.~\ref{subfig:a-match_140} and~\ref{subfig:b-match_140}: $Q = 3~{\rm GeV}$ and $\xB = 10^{-3}$ for the CM12 and SV13 LDME sets, respectively; Figs.~\ref{subfig:c-match_140} and~\ref{subfig:d-match_140}: $Q = 25~{\rm GeV}$ and $\xB = 0.035$ for the same sets. The hard scale is set as $\mu = \sqrt{M_\psi^2 + Q^2}$. The blue band corresponds to the FO result (valid at high $q_\T$), the green bands to the TMD one (valid at low $q_\T$), and the red bands to the matched one (valid for all $q_\T$). The vertical lines with arrows correspond to the $q_\T$ value at which the relative errors of the collinear (dark blue) and TMD (dark green) frameworks overshoot $15\%$. Details on the different TMD bands are provided in the main text.}
  \label{fig: cross-section matched (140)}
\end{figure}
\begin{figure}[t]
  \centering    
  \subfloat[\label{subfig:a-match_45}]{\includegraphics[width=.47\linewidth, keepaspectratio]{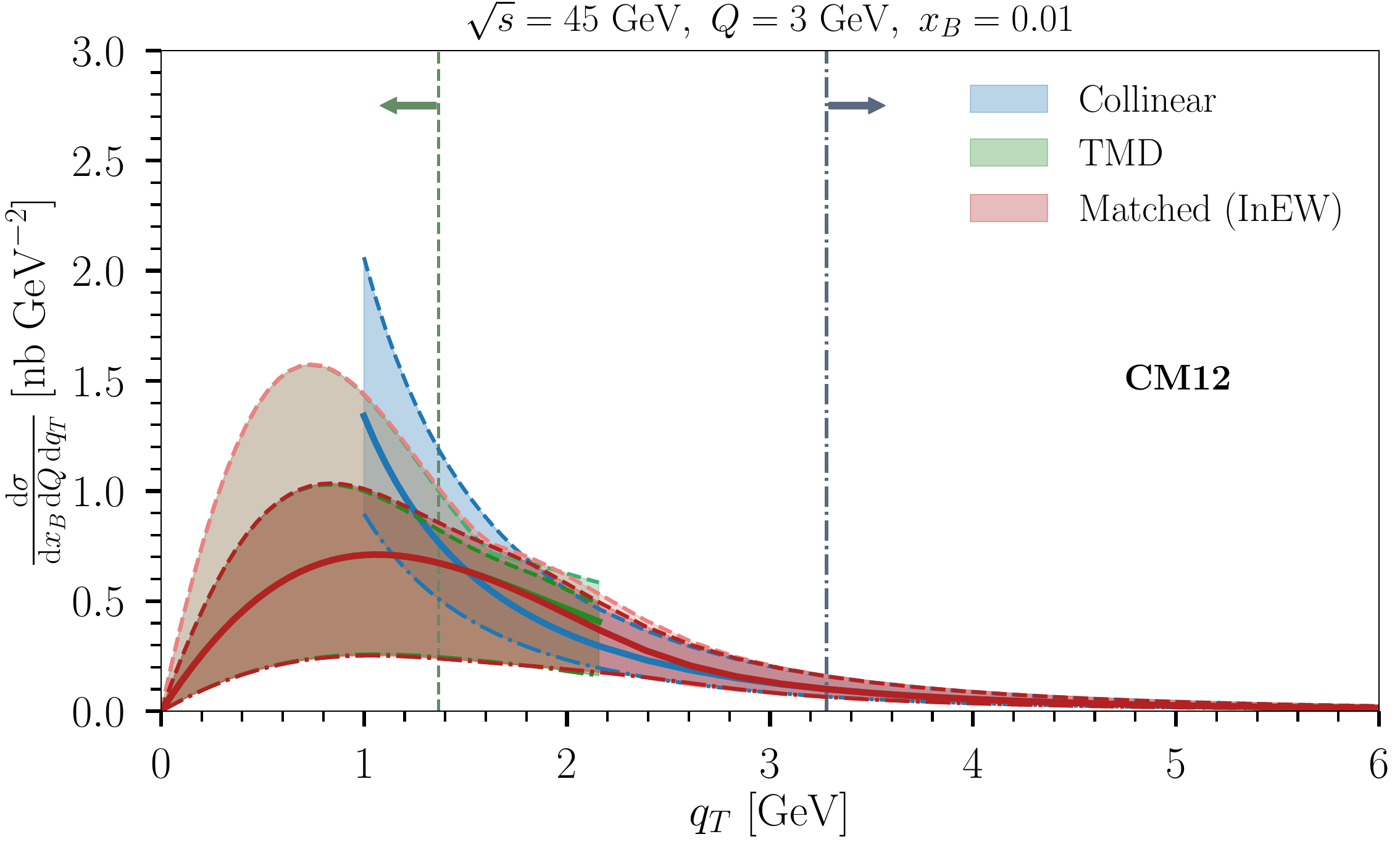}}
  \hfill
  \subfloat[\label{subfig:b-match_45}]{\includegraphics[width=.47\linewidth, keepaspectratio]{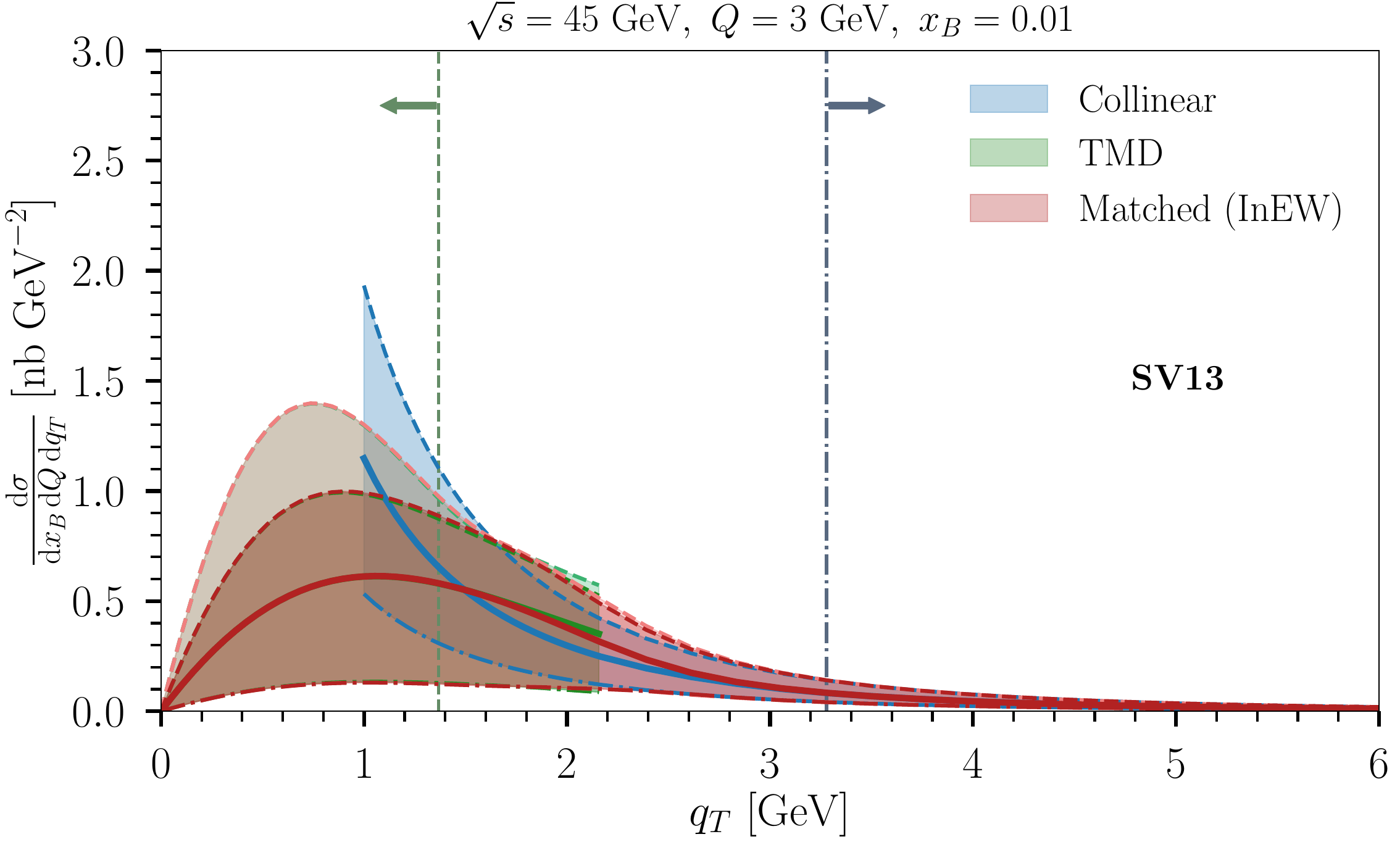}}\\
  \subfloat[\label{subfig:c-match_45}]{\includegraphics[width=.47\linewidth, keepaspectratio]{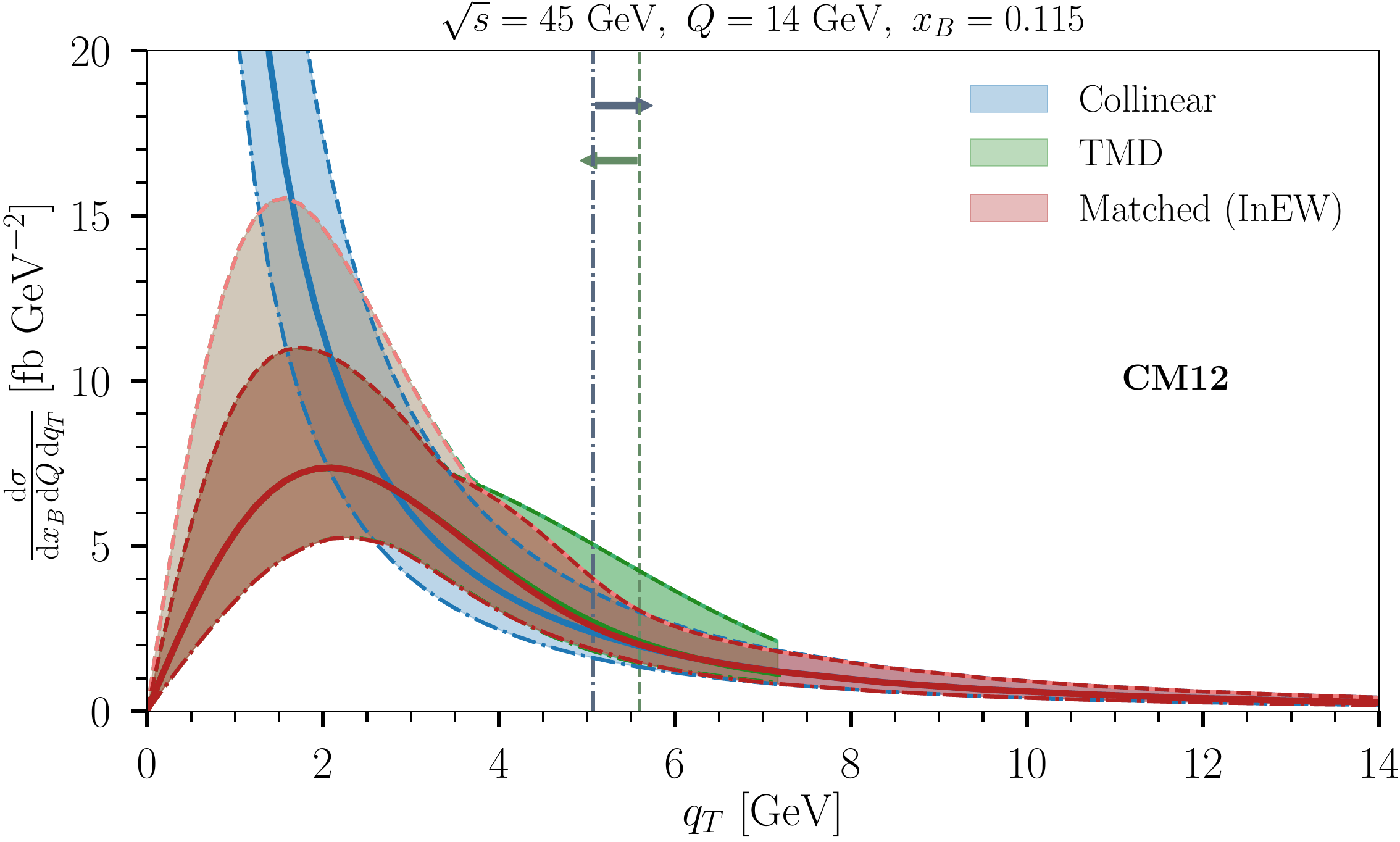}}
  \hfill
  \subfloat[\label{subfig:d-match_45}]{\includegraphics[width=.47\linewidth, keepaspectratio]{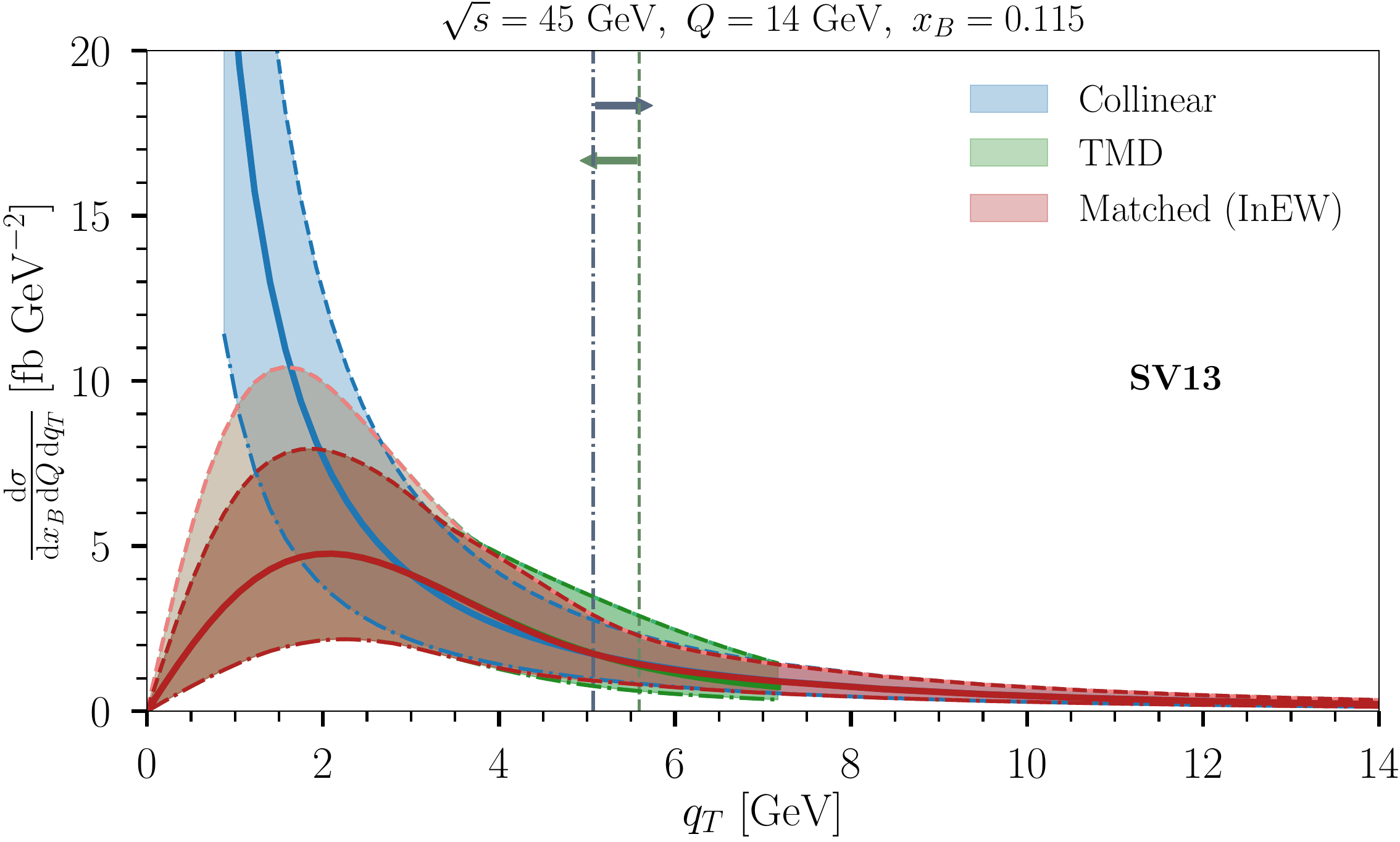}}
  \caption{\it Differential cross section with respect to $q_\T$ at the lowest nontrivial $\alpha_s$ order for $\sqrt{s} = 45~{\rm GeV}$. Figs.~\ref{subfig:a-match_140} and~\ref{subfig:b-match_140}: $Q = 3~{\rm GeV}$ and $\xB = 10^{-2}$ for the CM12 and SV13 LDME sets, respectively; Figs.~\ref{subfig:c-match_140} and~\ref{subfig:d-match_140}: $Q = 14~{\rm GeV}$ and $\xB = 0.115$ for the same sets. The hard scale is set as $\mu = \sqrt{M_\psi^2 + Q^2}$. The meaning of the vertical lines and different bands is the same as in Fig.~\ref{fig: cross-section matched (140)}.}
  \label{fig: cross-section matched (45)}
\end{figure}

In Figs.~\ref{fig: cross-section matched (140)} and~\ref{fig: cross-section matched (45)} we present the FO and TMD cross sections together with the matched curves obtained via the InEW method for two energies that are typically considered for the EIC, namely $\sqrt s = 140~{\rm GeV}$ and $\sqrt s = 45~{\rm GeV}$.
We consider two fixed values of $Q$ to identify both a low-$Q$ and high-$Q$ scenario, taking $Q = 3~{\rm GeV}$ and $Q = 25~{\rm GeV}$ for $\sqrt s = 140~{\rm GeV}$, and $Q = 3~{\rm GeV}$ and $Q = 14~{\rm GeV}$ for $\sqrt s = 45~{\rm GeV}$.
Note that by fixing $\xB$, $Q$ and $\sqrt s$, we also fix $y = \frac{Q^2}{\xB s}$ and $W_{\gamma p} = \sqrt{\frac{1 - \xB}{\xB}} Q$. 
For the parton distribution functions, we employ the NNPDF4.0LO set~\cite{NNPDF:2021njg}, while for the LDME set we use
~\cite{Chao:2012iv} denoted by CM12, 
and~\cite{Sharma:2012dy} denoted by SV13; 
the corresponding LDME values are reported in Tab.~\ref{tab: LDME values}. We note that, while the CM12 set is obtained at NLO accuracy, the SV13 set is extracted at LO precision.
Regarding the $g_\psi$ choice, all curves are obtained with $g_\psi = 0$, but more details on the dependence of the results on this nonperturbative parameter are given at the end of this section.
The TMD curves are shown up to ${q_\T = \mu/2}$ with $\mu = \sqrt{M_\psi^2 + Q^2}$, which is usually considered the border of applicability of the TMD factorization, whereas the collinear FO ones are shown down to ${q_\T = 1~{\rm GeV}}$, below which the collinear factorization is expected to no longer apply.

Each curve is presented with error bands. For the FO cross section, we have propagated, as independent errors, the LDME uncertainties included with each set, with the variation of the factorization scale by a factor of $2$. 
For the TMD cross section, the same LDME uncertainties are combined with the variation of the $A_{\rm NP}$ term in the nonperturbative Sudakov factor (see Eq.~\eqref{eq: nonperturbative Sudakov}), and the envelope error of $C_1$, $C_2$ variations by a factor of $2$, with $C_3 = C_1$.\footnote{In some studies, e.g.~\cite{Melis:2015ycg,Bor:2025ztq}, the case $C_3 \neq C_1$ is also considered, however, here we consider only the more conventional choice of $C_3 = C_1$, in order to keep the scale to which we evolve and the scale at which the PDFs are evaluated equal.}
More specifically, we consider both a $7$-points and a $9$-points variation.
In the first one, we additionally impose that ${0.5 < C_2/C_1 < 2}$, while in the second one, we have no constraints on the $C_2/C_1$ ratio. In the figures, the $7$-point variation band is the darker green one, whereas the $9$-point variation is the lighter green one.
For the approximated cross section in Eq.~\eqref{eq: cross section with InEW approx}, we have connected the $A_{NP}$ variation in the nonperturbative Sudakov factor and the scale variation of each TMD curve with the factorization scale variation of the collinear one, and combined them with the LDME error (now evaluated for the whole approximated cross section) and the error associated with the matching procedure in Eq.~\eqref{eq: intrinsic error InEW}, again considered as independent errors. 
This results in the bands associated with the matched curve, with the darker red constructed from the $7$-point variation band and the light red from the $9$-point variation one.

\begin{table}[t]
\centering
\resizebox{.7\columnwidth}{!}{
\begin{tabular}{|c|c|c|c|c|}
\toprule
\rule[-1.5ex]{0pt}{4.5ex} 
{\sc LDME set} & $\langle {\cal O}_\psi [^3 S_1^{(1)}] \rangle/{\rm GeV}^3$ & $\langle {\cal O}_\psi [^1 S_0^{(8)}] \rangle/{\rm GeV}^3$ & $\langle {\cal O}_\psi [^3 S_1^{(8)}] \rangle/{\rm GeV}^3$ & $\langle {\cal O}_\psi [^3 P_0^{(8)}] \rangle/{\rm GeV}^5$ \\ \midrule
\rule[-1.2ex]{0pt}{4.ex} \cite{Chao:2012iv} & $1.16$ & $0.0890\pm0.0098$ & $0.0030\pm0.00012$ & $(0.0056\pm0.0021)\, m_c^2$  \\  
\rule[-1.2ex]{0pt}{4.ex} \cite{Sharma:2012dy} & $1.2$ & $0.0180\pm0.0087$ & $0.0013\pm0.0013$ & $(0.0180\pm0.0087)\, m_c^2$  \\ 
\bottomrule
\end{tabular}
}
\caption{\it LDME values with the corresponding errors for each set, as taken from the corresponding reference.}
\label{tab: LDME values}
\end{table}

As can be seen in Fig.~\ref{fig: cross-section matched (45)}, at lower energies the cross section displays similar features as at higher energies, but with smaller magnitudes due to the higher $\xB$ (and thus light-cone fractions) explored. Moreover, the lower energy also further restricts the range of measurable $Q$ values. Hence, while we present both energies for completeness, we focus the discussion on higher energies, \textit{i.e.},\ on Fig.~\ref{fig: cross-section matched (140)}. 

We observe that for $Q = 3~{\rm GeV}$ (Figs.~\ref{subfig:a-match_140} and~\ref{subfig:b-match_140}), the transition from the TMD factorization expression to the collinear one happens in a region where both the theoretical errors (as defined in Eq.~\eqref{eq: scheme theoretical errors}) are larger than $15\%$ of the exact cross section. 
Nevertheless, a transition between the TMD and collinear predictions must occur and it is thus reasonable to assume that the matched curve can be evaluated by the same matching procedure discussed above, albeit with an intrinsic higher uncertainty. Regardless, we observe that the FO and TMD predictions still overlap, even if one considers the $7$-points variation solely, which is most likely underestimating the TMD error band. Of course, the result is also sensitive to the LDME values and uncertainties. In particular, we have observed that the SV13 set, due to the LDME uncertainties, allows negative cross section values at small $q_\T$ for certain combinations of $\xB$ and $Q$, and in particular for very low $\xB$ ($x$) and $Q$. We have also checked that the problem stays if one employs other PDF sets too, namely NNPDF4.0NLO~\cite{NNPDF:2021njg}, and MSHT20 both at LO and NLO~\cite{Bailey:2020ooq}.
This reflects the fact that the LDME set is insufficiently constrained, which can be improved by including data from the TMD region in their extraction. This problem is also observed for other sets, \textit{e.g.}, those reported in \cite{Butenschoen:2011yh}. 

In Figs.~\ref{subfig:c-match_140} and~\ref{subfig:d-match_140}, which correspond to $Q = 25~{\rm GeV}$, we observe a significant drop of the cross section, which is also, and mostly, related to the higher $\xB$. However, we expect that the EIC, with its high luminosity, will still be able to measure such events~\cite{AbdulKhalek:2021gbh}. The main reason to focus on the higher $Q$ region is that it enhances the effect of $B_{\rm CO}$ (including $B_{ep}$) in the Sudakov factor.  
Also, the higher $Q$ extends significantly the TMD region, with the same threshold of $15\%$ on the relative errors $\Delta W/\sigma$ and $\Delta {\rm FO}/\sigma$ now identifying a $q_\T$ interval where both factorizations are considered applicable.
We observe that the curves for the two factorizations indeed overlap in this region, independently from the LDME set employed. This is also observed for $Q=14~{\rm GeV}$ too (Figs.~\ref{subfig:c-match_45} and~\ref{subfig:d-match_45}), where the overlapping region is expected to be much smaller.
Furthermore, we note that Figs.~\ref{subfig:c-match_140} and~\ref{subfig:d-match_140} display hints of a double-peak structure of the matched curve, especially for the $7$-point variation. This is related to the different behaviors of the TMD and collinear curves as a function of $Q$, where the former decreases faster than the latter, leading to the presence of a double peak in the matched cross section for sufficiently high $Q$ values.

\begin{figure}[t]
  \centering    
  \subfloat[\label{subfig:a-match140-cm}]{\includegraphics[width=.47\linewidth, keepaspectratio]{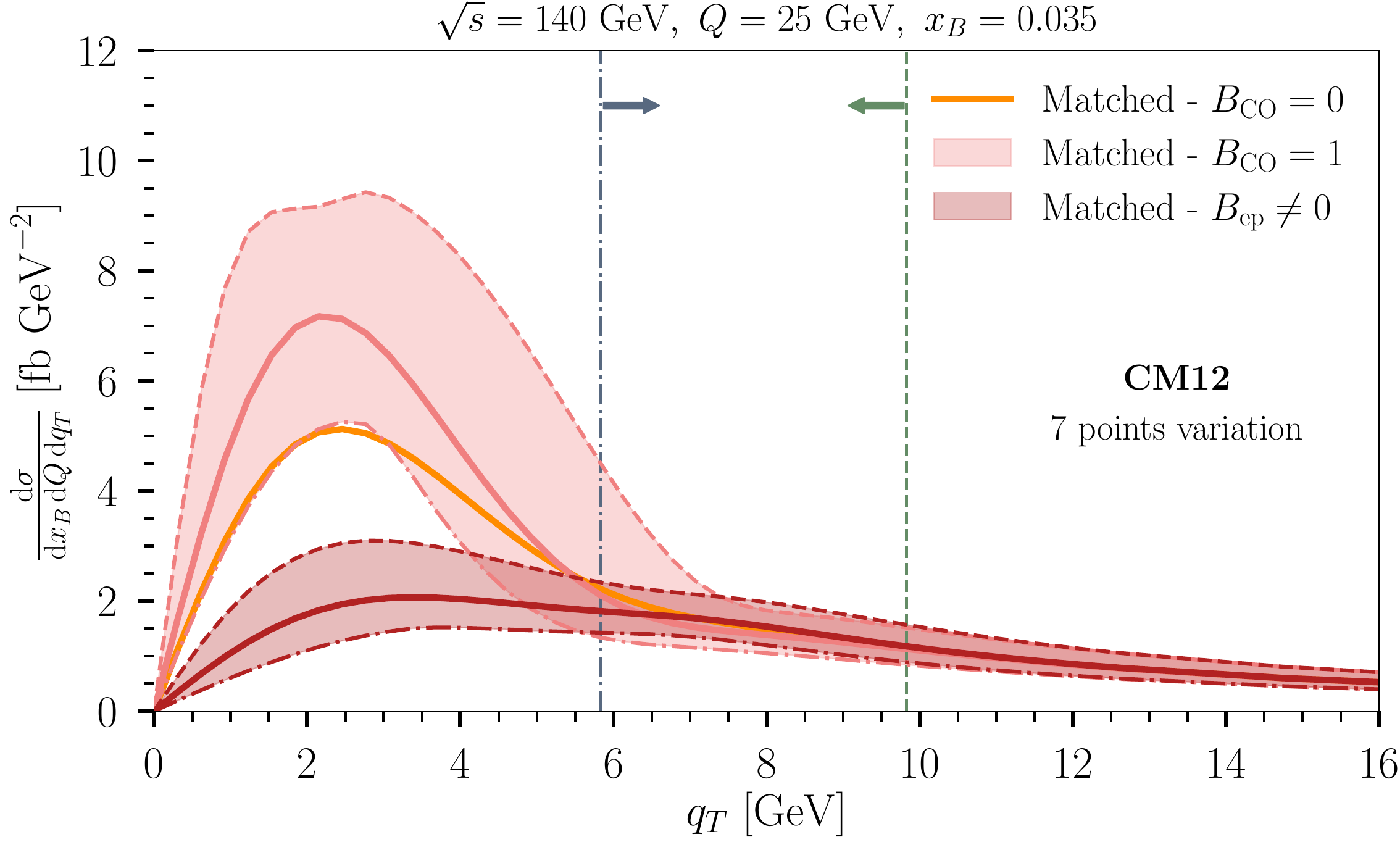}}
  \hfill
  \subfloat[\label{subfig:b-match140-sv}]{\includegraphics[width=.47\linewidth, keepaspectratio]{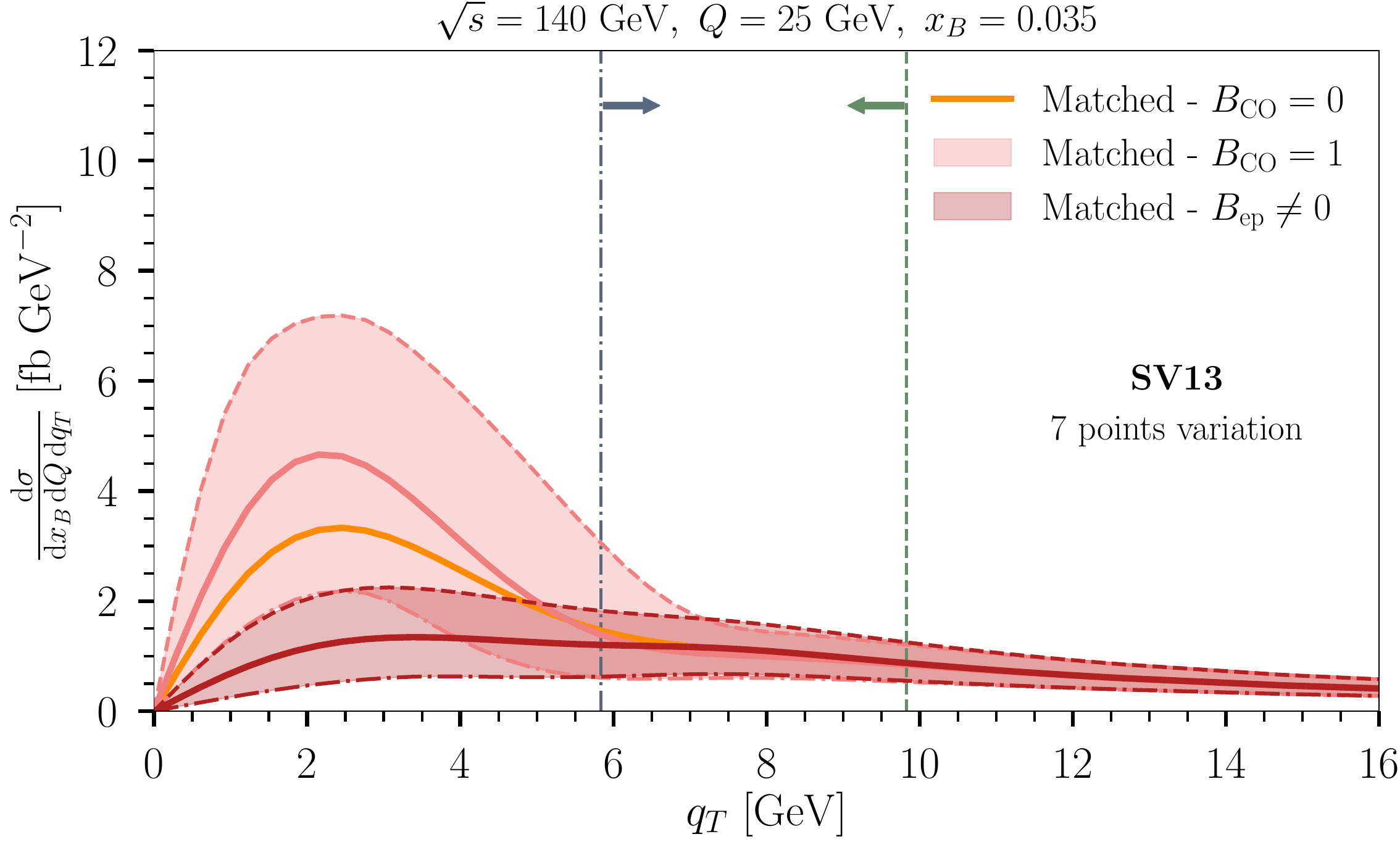}}\\
  \subfloat[\label{subfig:c-match45-cm}]{\includegraphics[width=.47\linewidth, keepaspectratio]{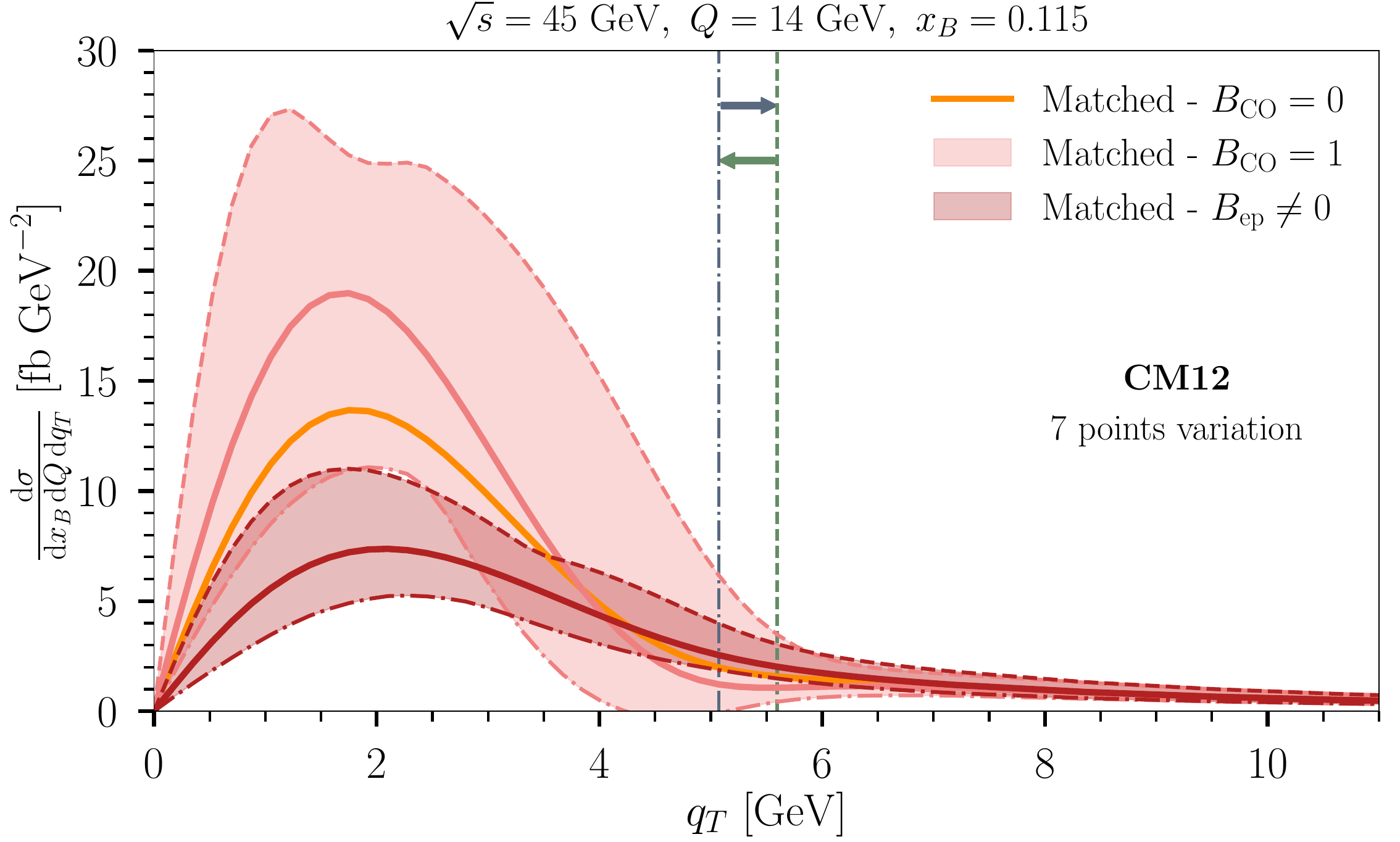}}
  \hfill
  \subfloat[\label{subfig:d-match45-sv}]{\includegraphics[width=.47\linewidth, keepaspectratio]{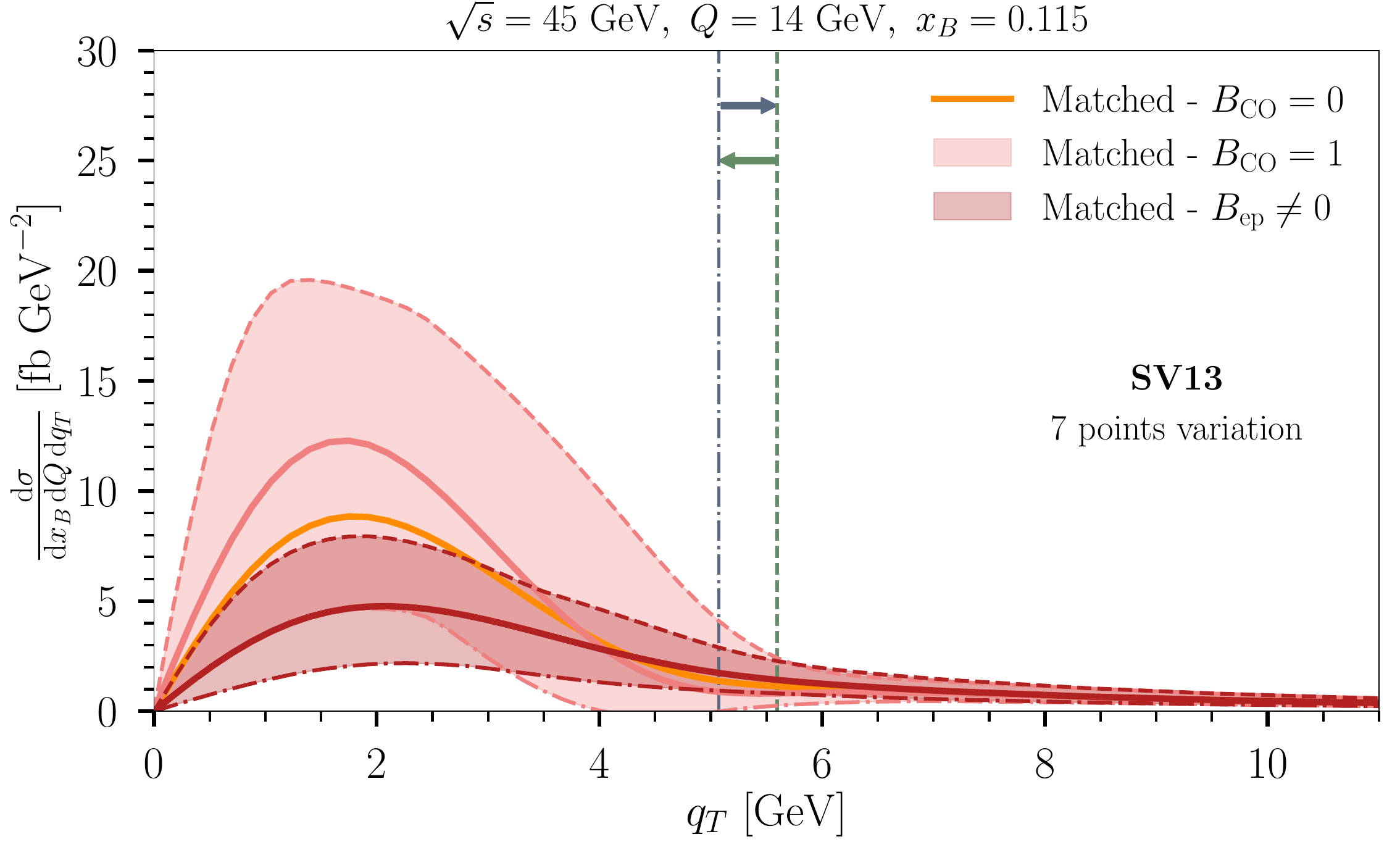}}\\
  \caption{\it Dependence of the matched differential cross section for different ${B_{\rm CO} = B_\psi + B_{ep}}$ choices, with $B_{\rm CO} = 1$ from Ref.~\cite{Echevarria:2024idp} and $B_{ep} \neq 0$ from Ref.~\cite{Boer:2023zit}. Upper panels: $\sqrt s = 140~{\rm GeV}$ for $Q=25~{\rm GeV}$ and $\xB = 0.035$  using CM12 (\ref{subfig:a-match140-cm}) and SV13 (\ref{subfig:b-match140-sv}). Lower panels: $\sqrt s = 45~{\rm GeV}$ for $Q=14~{\rm GeV}$ and $\xB = 0.115$ using CM12 (\ref{subfig:c-match45-cm}) and SV13 (\ref{subfig:d-match45-sv}). For clarity, we do not show the error band for the case $B_{\rm CO} = 0$, but only its central value, namely ${C_1 = C_2 = C_3 = 1}$.}
  \label{fig: TMD-ShF investigation}
\end{figure}

Next we discuss the observation that at high $Q$ the effects of the perturbative tail of the TMD-ShF become significant.
In Fig.~\ref{fig: TMD-ShF investigation} we present the dependence of the matched curve on the $B_{\rm CO}$ choice, taking $\sqrt s = 140~{\rm GeV}$ and $Q = 25~{\rm GeV}$, and $\sqrt s = 45~{\rm GeV}$ and $Q = 14~{\rm GeV}$. We compare the no-TMD-ShF case ($B_{\rm CO} = 0$) for its central value without error band with the results obtained using the $B$ corresponding to the choice of Ref.~\cite{Echevarria:2024idp} ($B_{\rm CO} = 1$ with $B_{ep} = 0$) and of \cite{Boer:2023zit} ($B_{ep} \neq 0$).
For clarity, we only include the error bands obtained from the $7$-points variation for the TMD factorization expression.
The figures clearly display the sensitivity of the cross section to TMD effects in the final state. In particular, TMD contributions driven by $B_\psi$ enhance the curve expectation at low $q_\T$, whilst keeping it peaked at more or less the same $q_\T$, whereas $B_{ep}$ has the opposite effect, causing a suppression of the curve and a potential shift of the peak position to higher $q_\T$ values.
Of course, the difference between the $B_{\rm CO} = 1$ and $B_{ep} \neq 0$ cases is LDME-dependent, as the LDMEs, besides having different uncertainties, also determine the magnitude of the cross section. Hence, LDMEs that lead to larger (smaller) cross sections at low $q_\T$ will enhance (suppress) the difference between the two cases, as also seen in Fig.~\ref{fig: TMD-ShF investigation}. 
Given these large differences in magnitude, we expect that besides providing improved determinations of LDMEs, future EIC data at low $q_\T$ may be precise enough to discern nontrivial process-dependent effects in the soft factor of the process.
Additionally, we observe that the $B_{\rm CO}=1$ case (and to some extent $B_{\rm CO} = 0$ too) can lead to negative cross sections within the TMD framework for certain combinations of $\mu$ and $\xB$. In particular, this happens for $\Lambda_{\rm QCD} \ll q_\T \lesssim \mu/2$ at sufficiently high scales (for instance at the scales $Q = 14~{\rm GeV}$ and $Q = 25~{\rm GeV}$ we consider here). Although not always visible in the matched curves we present, for $Q=14~{\rm GeV}$ the effect of a negative TMD cross section can be observed in Fig.~\ref{fig: TMD-ShF investigation}, with the band allowing negative values for $B_{\rm CO} = 1$, and not for $B_{ep}\neq 0$.
Since this behavior is primarily driven by the convolution of the TMD quantities, it is independent of the LDME set used. Thus, we interpret it as a sign of missing contributions.

Finally, we conclude this section by discussing the effect of a nontrivial $g_\psi$ dependence in the nonperturbative Sudakov factor, Eq.~\eqref{eq: nonperturbative Sudakov}. We have checked that, as expected, the position of the peak of the distribution depends on the exact $g_\psi$ value. However, the variation is much smaller compared to the band shown here, and nonperturbative effects are then washed out by the large error bands from the TMD scale variation.

\section{The \texorpdfstring{$\cos(2\phi_\psi)$}{cos(\textit{2φ})} asymmetry in unpolarized collisions}
\label{sec: cos2phi}
We now discuss the $\cos(2\phi_\psi)$ asymmetry. The average asymmetry is defined as
\begin{equation}
    \langle \cos(2\phi_\psi) \rangle = 2\, \frac{\int \d \sigma\, \cos(2\phi_\psi)\, \d\phi_\psi}{\int \d \sigma\, \d\phi_\psi} = \frac{(1 - y)\, F_{UU}^{\cos 2\phi_\psi}}{(2 - 2 y + y^2)\, F_{UUT} + 4\,(1 - y)\, F_{UUL}}\ ,
\label{eq: cos2phi average asymmetry}
\end{equation}
where we have used the shorthand notation $\d \sigma \equiv \frac{\d \sigma}{ \d z\, \d \xB\, \d Q\, \d q_\T^2 \d \phi_\psi}$, and we have re-expressed the asymmetry in terms of structure functions in the last step. By definition, we have that $|\langle \cos(2\phi_\psi) \rangle| \leq 1$. Note that below, we will also refer to the denominator of Eq.~\eqref{eq: cos2phi average asymmetry} as $\d \sigma_{\rm iso}$ for convenience, since it corresponds to the isotropic cross section discussed in the previous section. Moreover, we remark that within the TMD factorization, $F_{UU}^{\cos 2\phi_\psi}$ originates from the linearly polarized TMD distribution $h_1^{\perp g}$, of which the sign is unknown and may actually be a function of $x$ and $k_T$.\footnote{In the following we will use the perturbative expansion in $b_\T$ space of its Fourier transform $\tilde h_1^{\perp g}(x, b_\T)$ from \cite{Bor:2022fga}, as we are considering the same correlator as in that reference, albeit non explicitly reported here. Note that although different correlator definitions in terms of TMD distributions are used in the literature and will modify the prefactor of the expansion (in particular in sign and/or factors of $2$), this should of course not affect the final result.} On the other hand, in collinear factorization the same numerator depends on the unpolarized distribution only, whose sign is defined.
The transition from the collinear to TMD factorization happens at the cross section level, rather than the asymmetry. Hence, the cross section in the numerator and denominator of Eq.~\eqref{eq: cos2phi average asymmetry} are expanded according to Eq.~\eqref{eq: cross section with InEW approx} separately, which leads to
\begin{align}
    \langle \cos(2\phi_\psi) \rangle & 
    \approx \frac{(1 - y)\, \Big(\omega_1\, F_{UU}^{\cos 2\phi_\psi}\big|_{\rm TMD} + \omega_2\, F_{UU}^{\cos 2\phi_\psi}\big|_{\rm FO} \Big)}{\omega_1\, \d \sigma_{\rm iso}\big|_{\rm TMD} + \omega_2\, \d \sigma_{\rm iso}\big|_{\rm FO}} \nonumber
    \\
    & = 
    \omega_1 \left(1 - \frac{ (1 - \omega_1)\, \d \sigma_{\rm iso}\big|_{\rm TMD} - \omega_2\, \d \sigma_{\rm iso}\big|_{\rm FO} }{ \omega_1\, \d \sigma_{\rm iso}\big|_{\rm TMD} + \omega_2\, \d \sigma_{\rm iso}\big|_{\rm FO} }\right) \langle \cos(2\phi_\psi) \rangle \big|_{\rm TMD} \nonumber \\
    & \phantom{=}  + \omega_2 \left( 1 - \frac{ (1 - \omega_2)\, \d \sigma_{\rm iso}\big|_{\rm FO} - \omega_1\, \d \sigma_{\rm iso}\big|_{\rm TMD} }{ \omega_1\, \d \sigma_{\rm iso}\big|_{\rm TMD} + \omega_2\, \d \sigma_{\rm iso}\big|_{\rm FO} } \right) \langle \cos(2\phi_\psi) \rangle \big|_{\rm FO}\ .
\label{eq: cos2phi matched}
\end{align}
The exact $\langle \cos(2\phi_\psi) \rangle$ is then approximated by the one evaluated within TMD and collinear factorization for $\omega_2 \to 0$ and $\omega_1 \to 0$, respectively. For intermediate $q_\T$, where $\omega_1$ and $\omega_2$ are comparable to each other, we need to take nonlinear terms in $\omega_1$ and $\omega_2$ into account in the approximated $\langle \cos(2\phi_\psi) \rangle$. Although one might expect that $\langle \cos2\phi_\psi \rangle \approx \omega_1 \langle \cos2\phi_\psi \rangle\big|_{\rm TMD} + \omega_2 \langle \cos2\phi_\psi \rangle\big|_{\rm FO}$ would be a sufficient estimator of the size of asymmetry at all $q_\T$ values, and it actually is for the majority of the kinematic region, the other terms cannot be neglected in general, as they can lead to sizable shifts (up to $10\%$). Hence, the complete approximated form in Eq.~\eqref{eq: cos2phi matched} will be used henceforth.

Some care must be taken when propagating the errors as well since the variations are also evaluated at the cross section level, and one cannot directly combine the TMD and FO bands.  Instead, one should take the complete expression in Eq.~\eqref{eq: cos2phi matched} and evaluate TMD variations while keeping the FO curve at its central value and vice versa. The two contributions can then be combined as independent errors, together with the LDME variation, that once again applies to the whole expression.

\begin{figure}[t]
  \centering    
  \subfloat[\label{subfig:a-asy140-pos}]{\includegraphics[width=.33\linewidth, keepaspectratio]{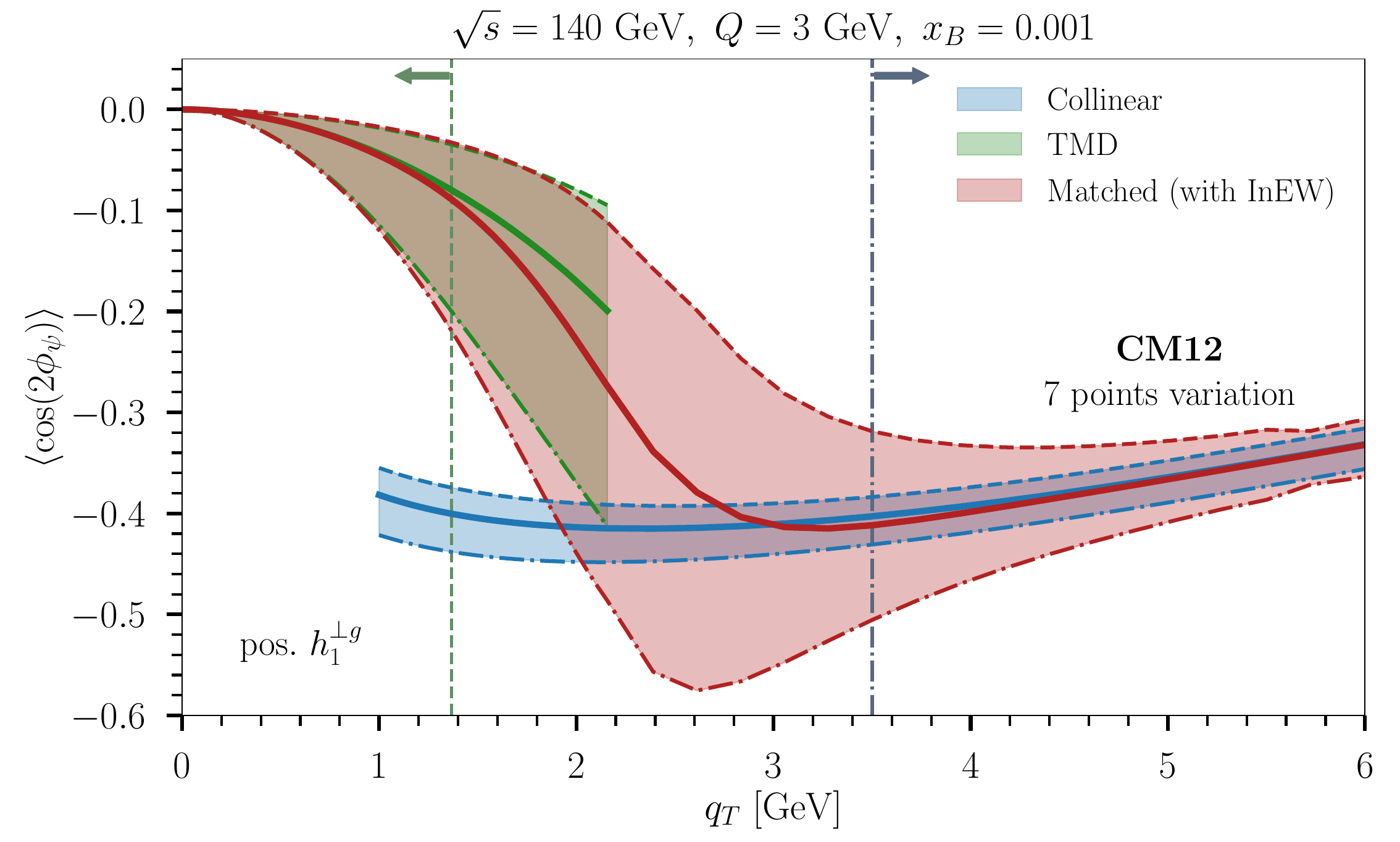}}
  \hfill
  \subfloat[\label{subfig:b-asy140-pos}]{\includegraphics[width=.33\linewidth, keepaspectratio]{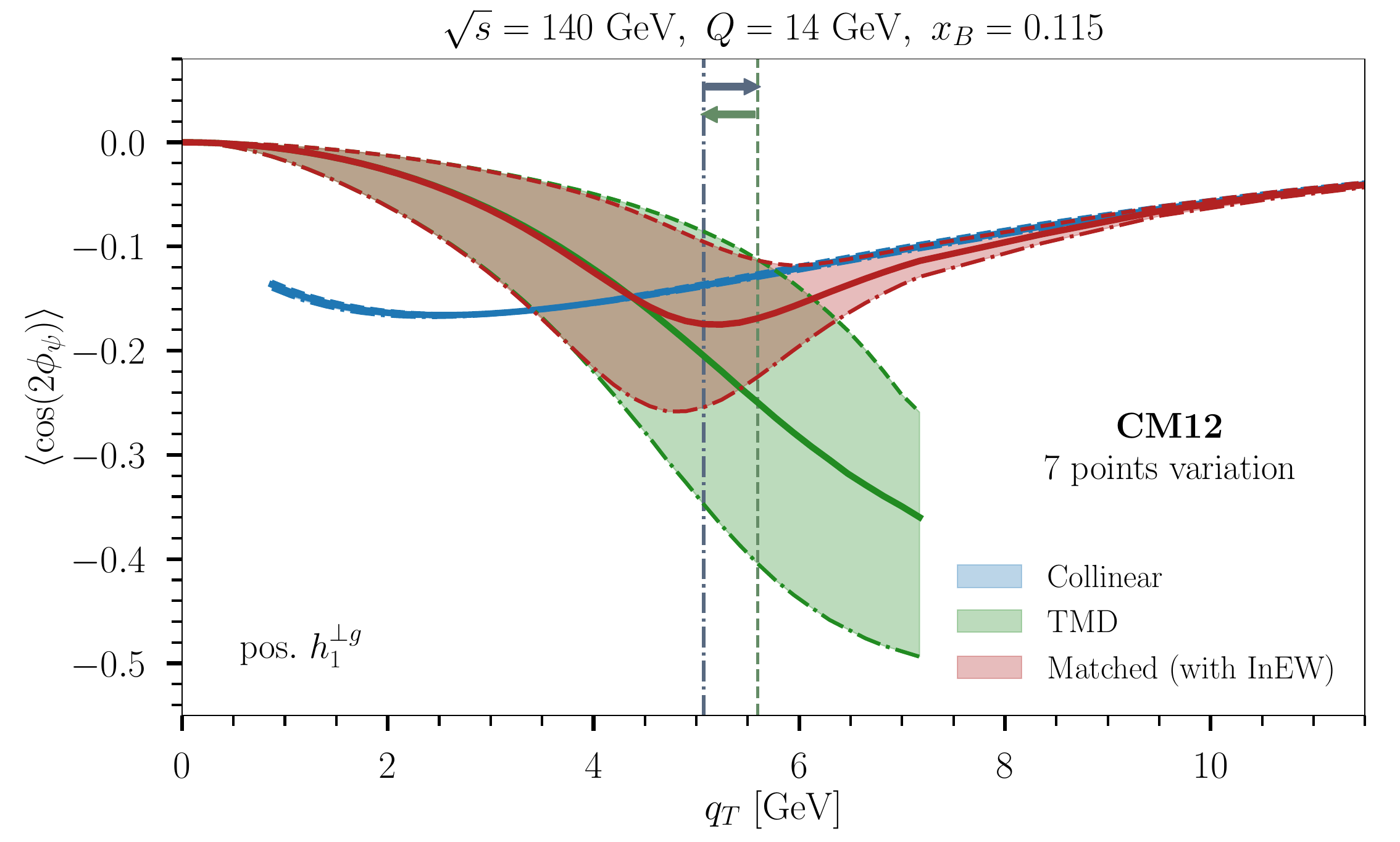}}\hfill
  \subfloat[\label{subfig:c-asy140-pos}]{\includegraphics[width=.33\linewidth, keepaspectratio]{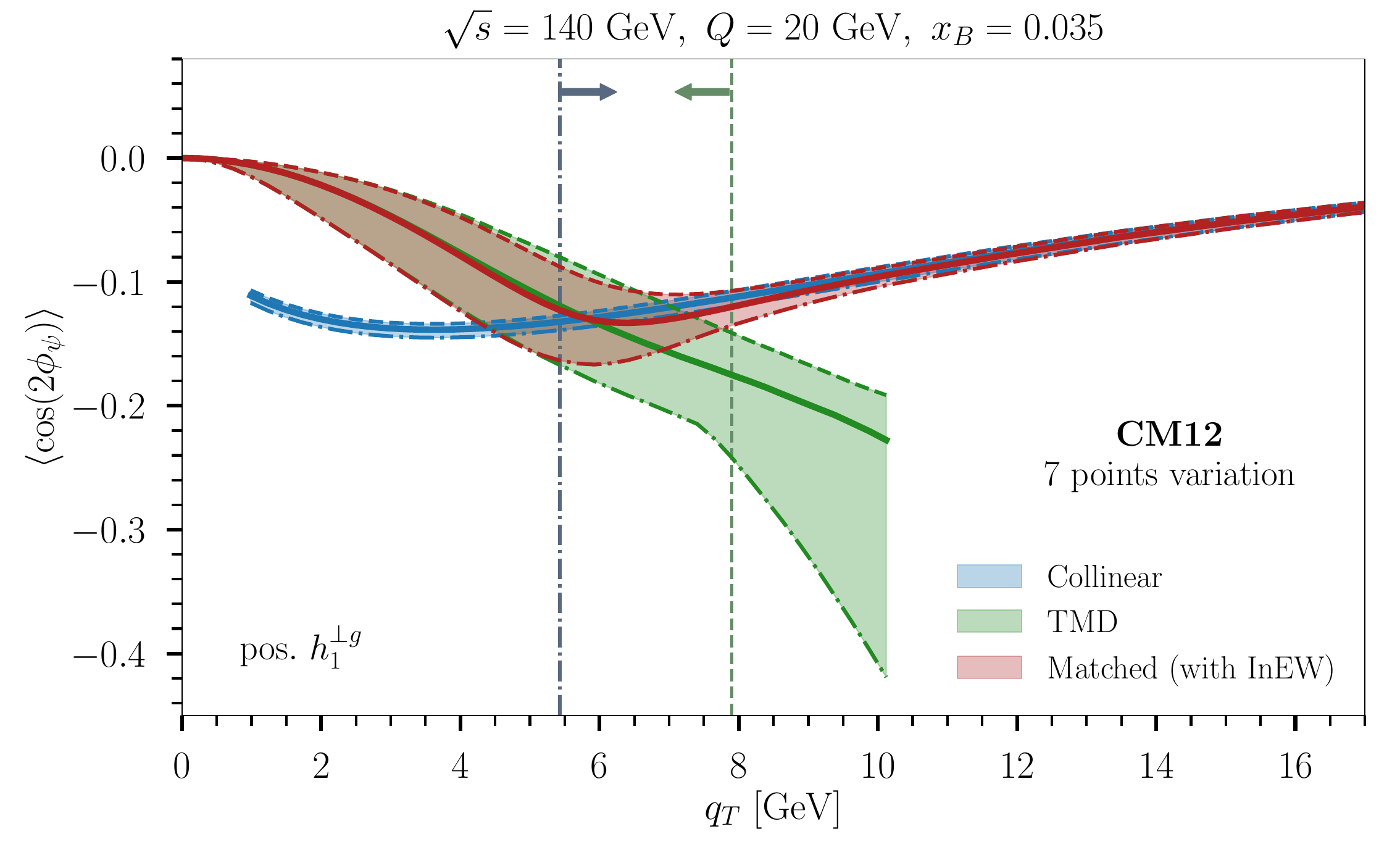}}\\
  \subfloat[\label{subfig:d-asy140-pos}]{\includegraphics[width=.33\linewidth, keepaspectratio]{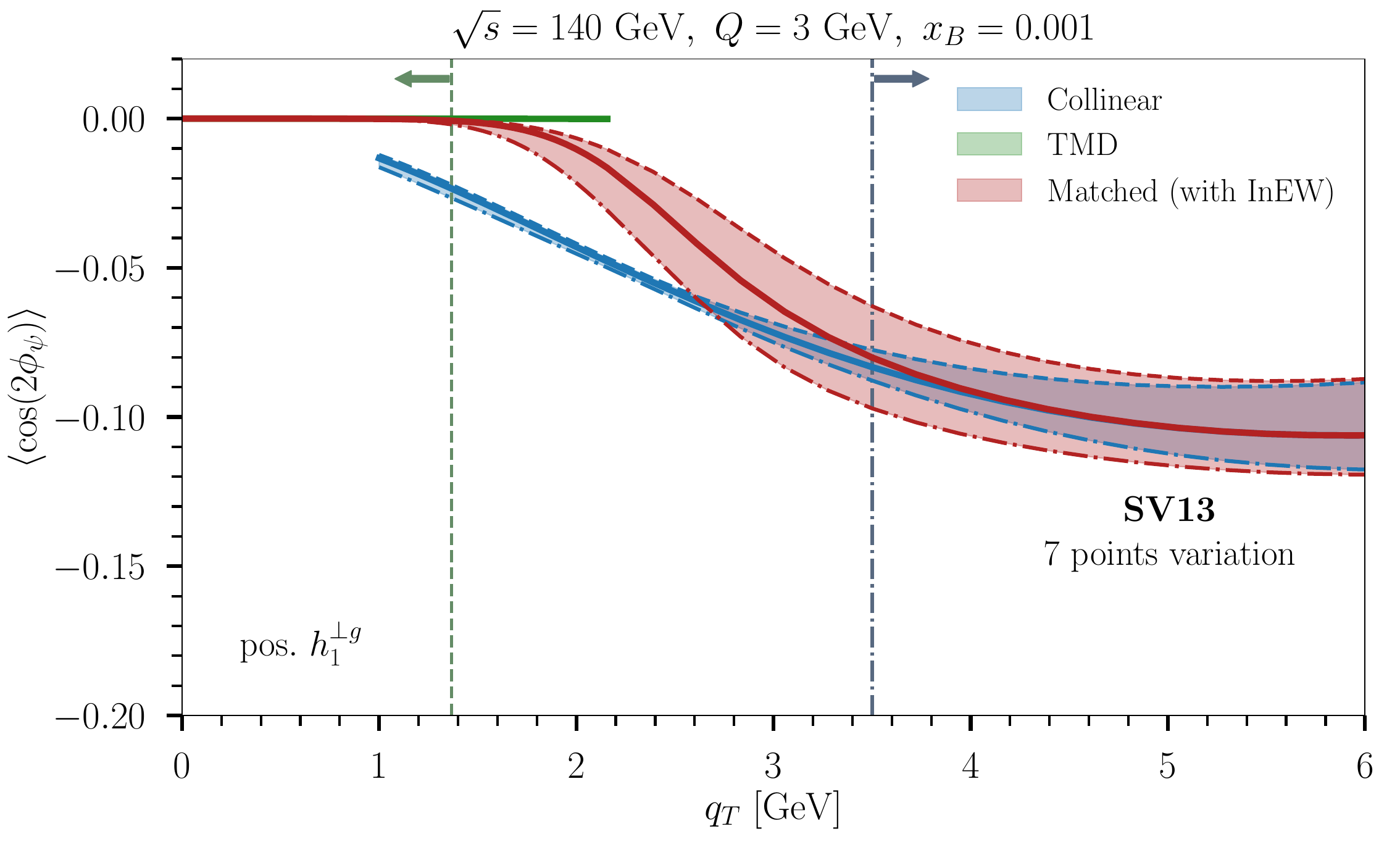}}
  \hfill
  \subfloat[\label{subfig:e-asy140-pos}]{\includegraphics[width=.33\linewidth, keepaspectratio]{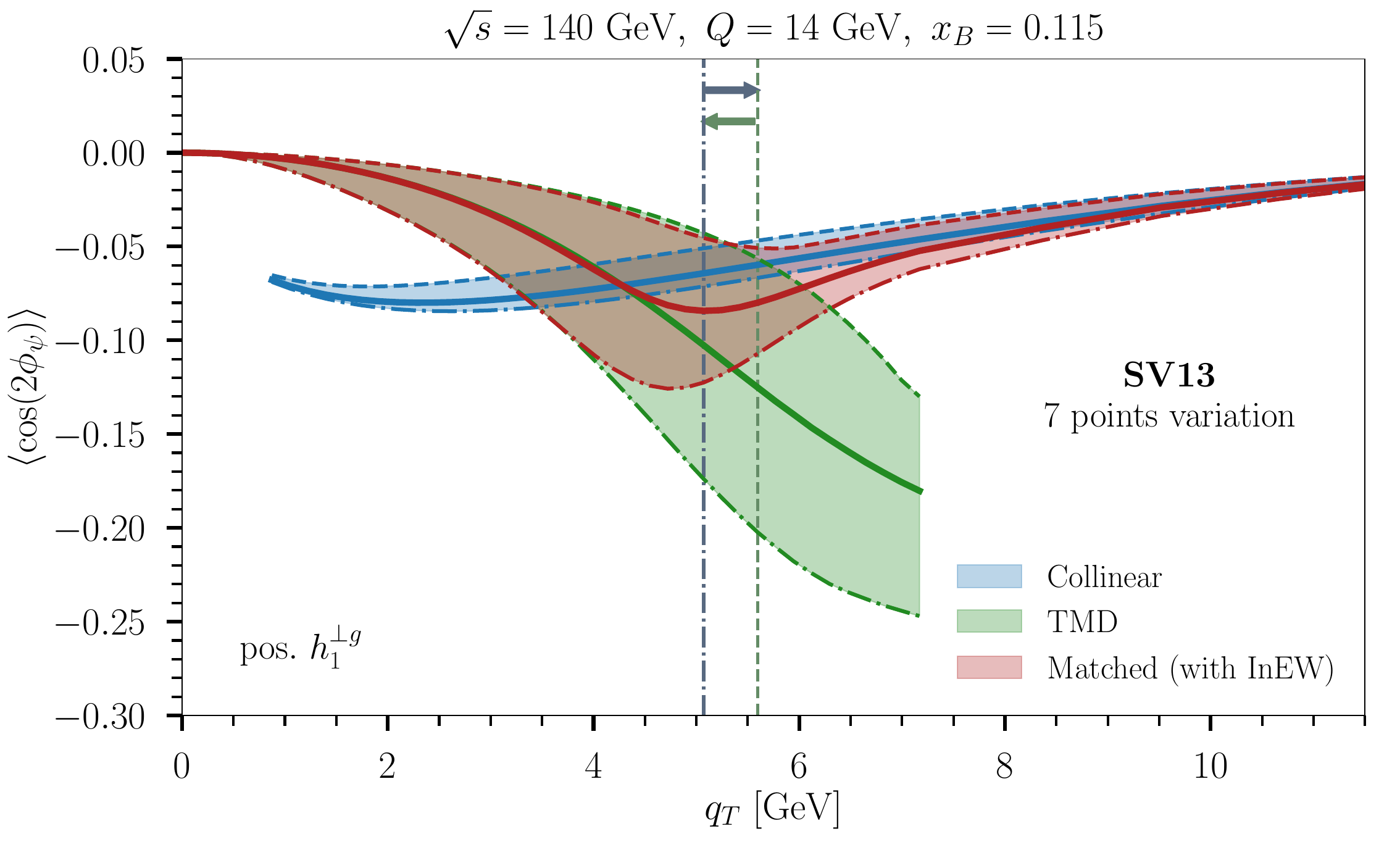}}\hfill
  \subfloat[\label{subfig:f-asy140-pos}]{\includegraphics[width=.33\linewidth, keepaspectratio]{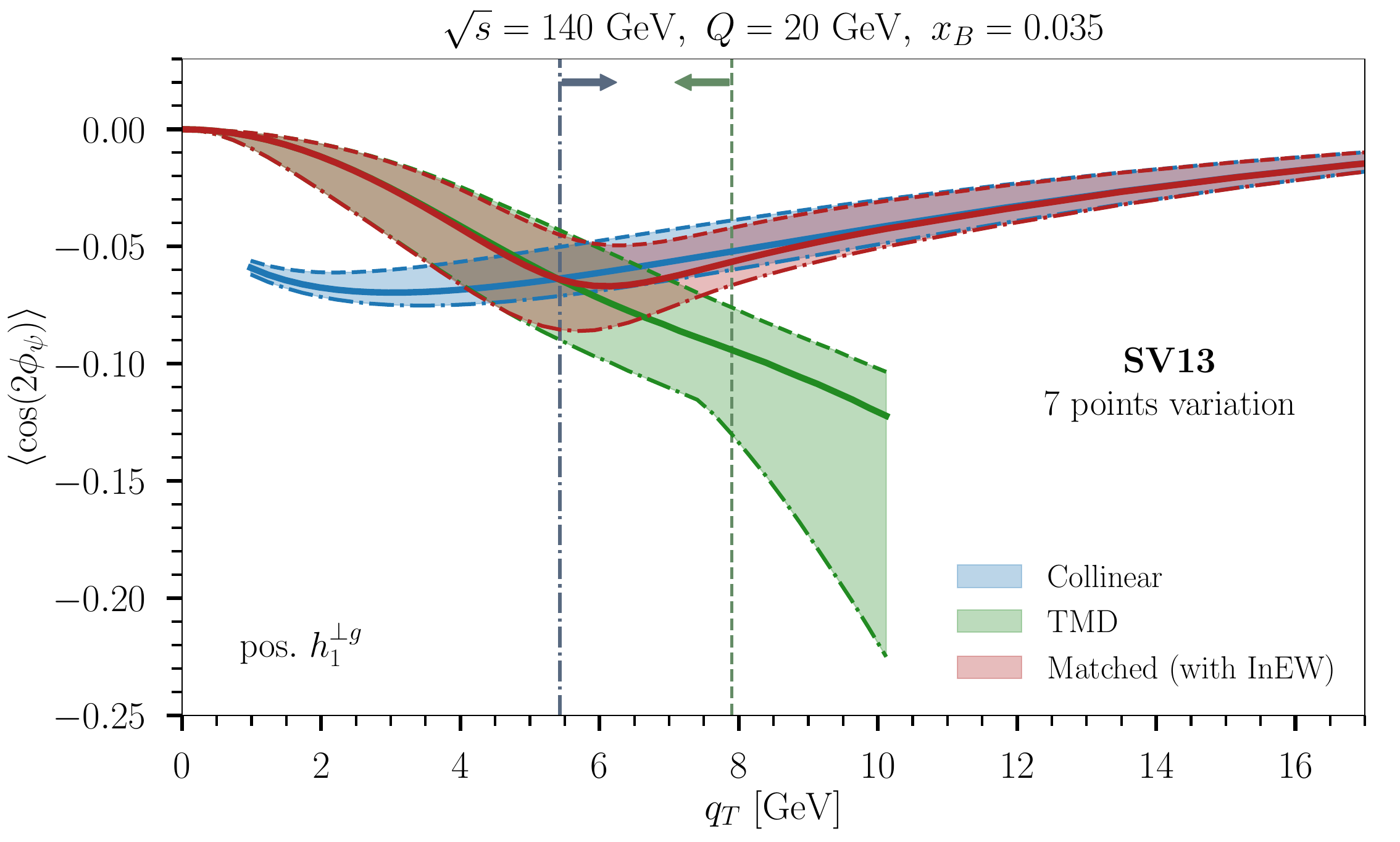}}
  \caption{\it Dependence of the matched $\langle \cos(2\phi_\psi) \rangle$ asymmetry on $q_\T$, at $\sqrt s = 140~{\rm GeV}$ and fixed kinematics: (a) and (d) $Q = 3~{\rm GeV}$ and $\xB = 10^{-3}$; (b) and (e) $Q = 14~{\rm GeV}$ and $\xB = 0.115$ ; (c) and (f) $Q = 20~{\rm GeV}$ and $\xB = 0.035$. Upper panels are evaluated using CM12, whereas SV12 is used for the lower ones. The factorization scale is taken as $\mu = \sqrt{M_\psi^2 + Q^2}$. For the TMD curve we have chosen the positive sign of the $h_1^{\perp g}$ function. For the explanation of the vertical lines see the caption of Fig.~\ref{fig: cross-section matched (140)} and for details on the evaluation of each band we refer to the text.}
  \label{fig: matched asymmetries (140) - pos}
\end{figure}

\begin{figure}[t]
  \centering    
  \subfloat[\label{subfig:a-asy140-neg}]{\includegraphics[width=.33\linewidth, keepaspectratio]{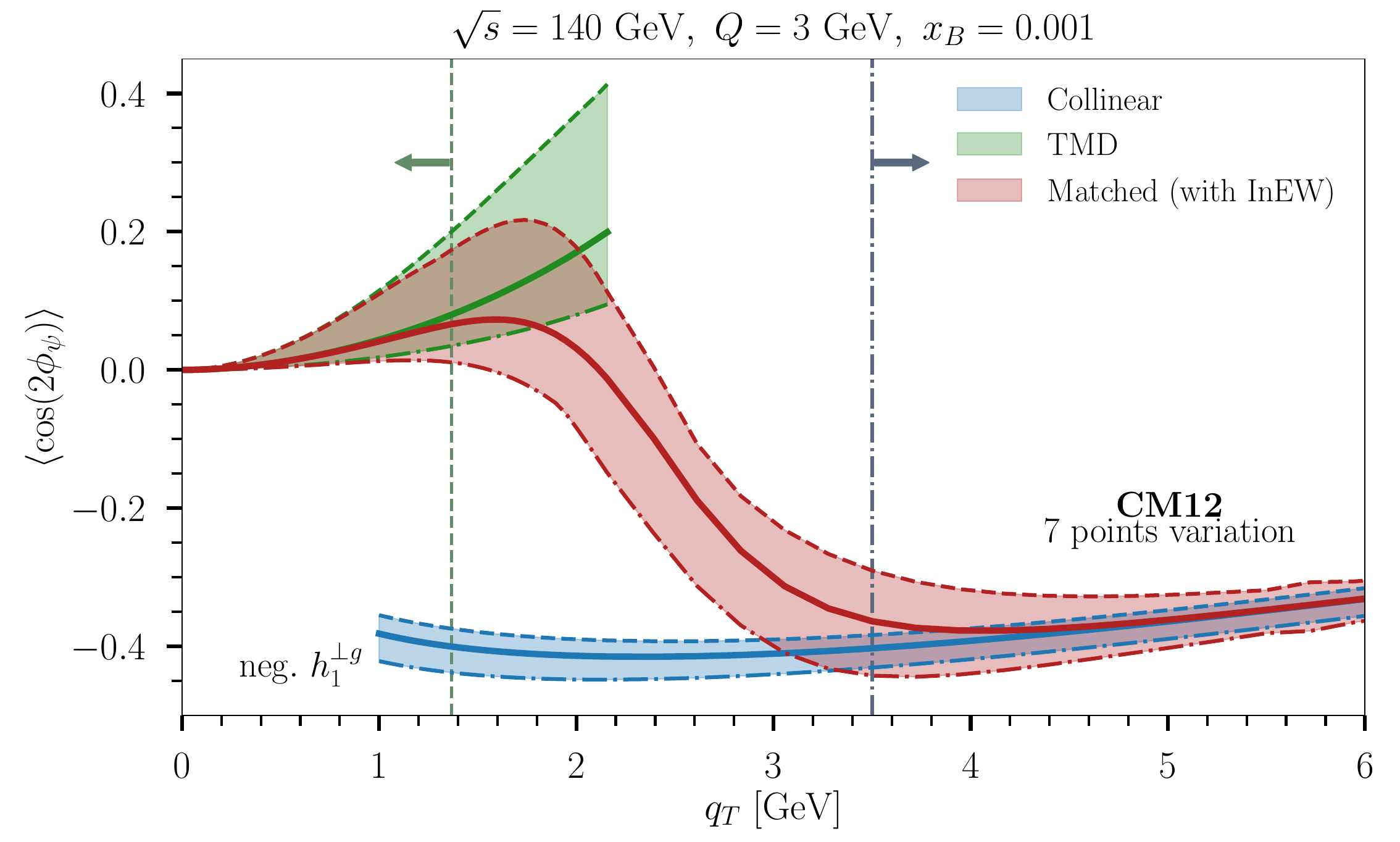}}
  \hfill
  \subfloat[\label{subfig:b-asy140-neg}]{\includegraphics[width=.33\linewidth, keepaspectratio]{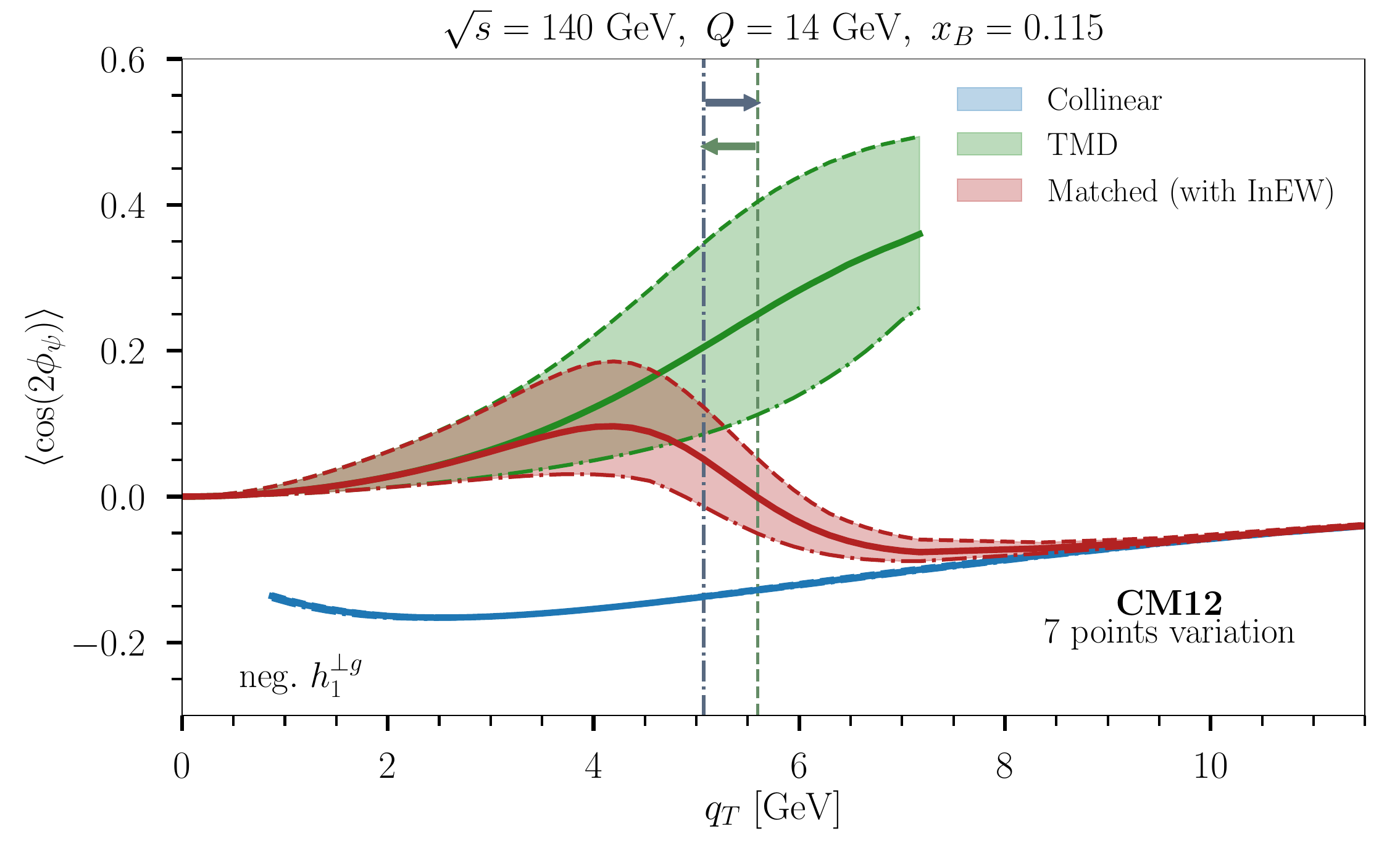}}\hfill
  \subfloat[\label{subfig:c-asy140-neg}]{\includegraphics[width=.33\linewidth, keepaspectratio]{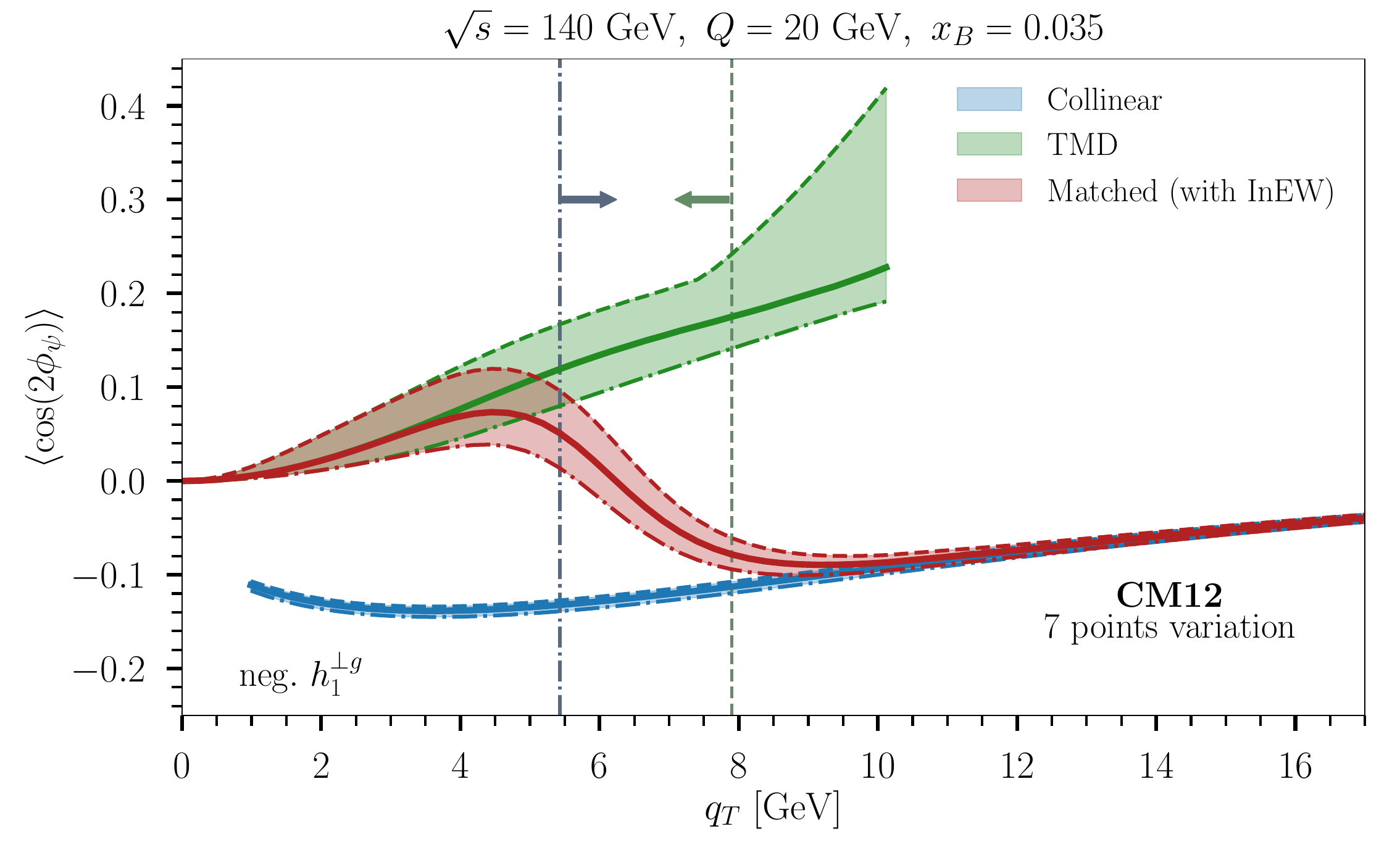}}\\
  \subfloat[\label{subfig:d-asy140-neg}]{\includegraphics[width=.33\linewidth, keepaspectratio]{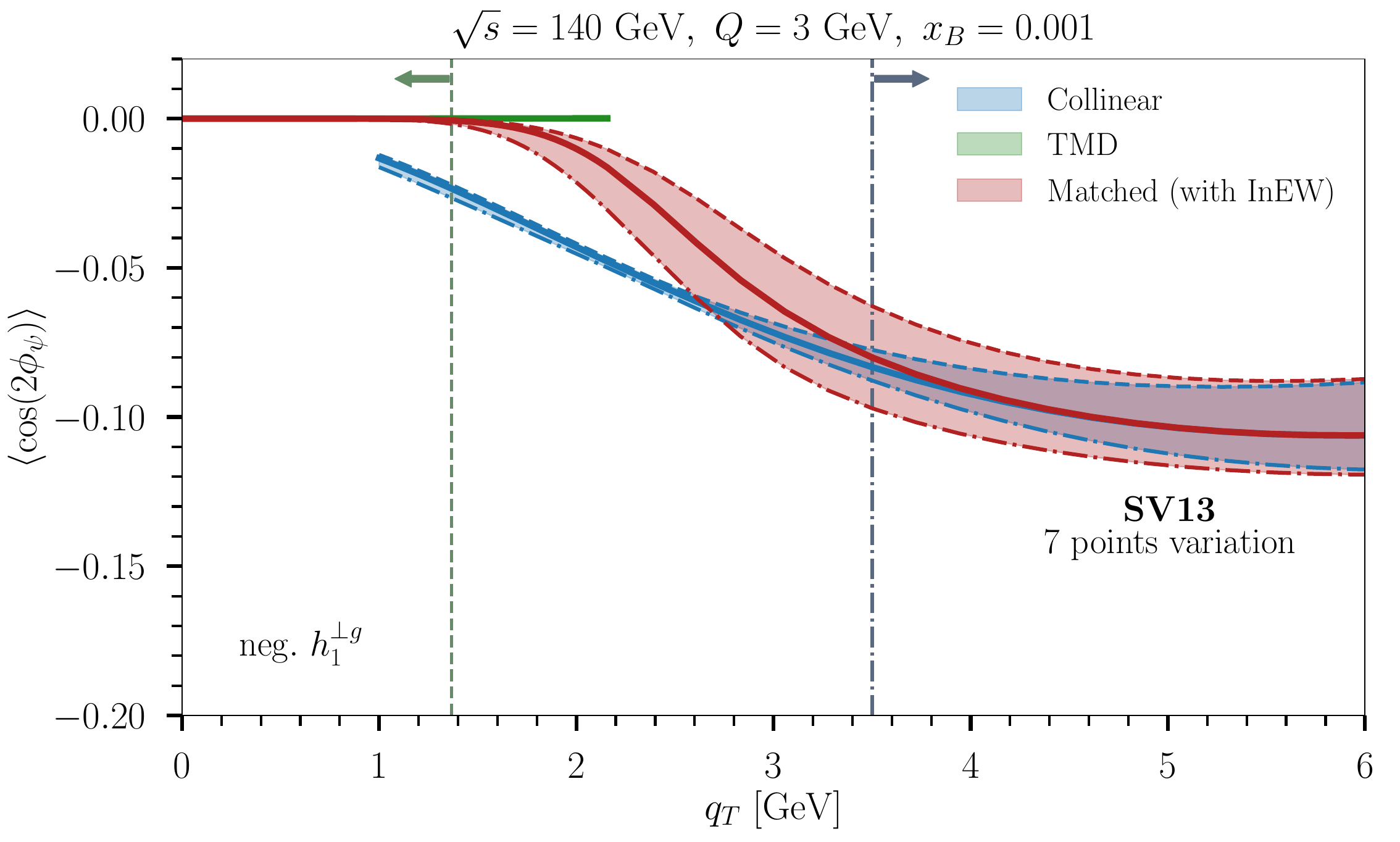}}
  \hfill
  \subfloat[\label{subfig:e-asy140-neg}]{\includegraphics[width=.33\linewidth, keepaspectratio]{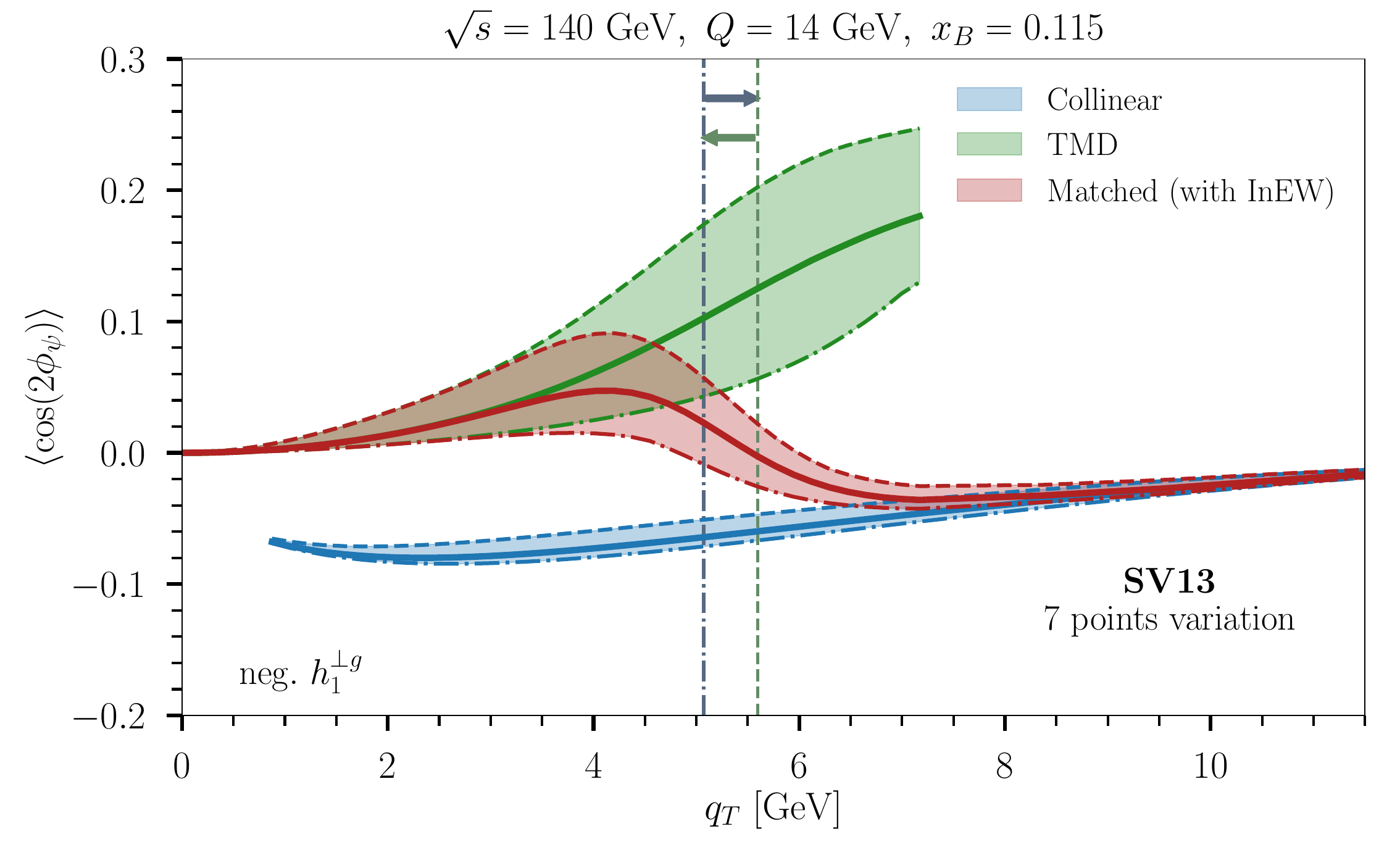}}\hfill
  \subfloat[\label{subfig:f-asy140-neg}]{\includegraphics[width=.33\linewidth, keepaspectratio]{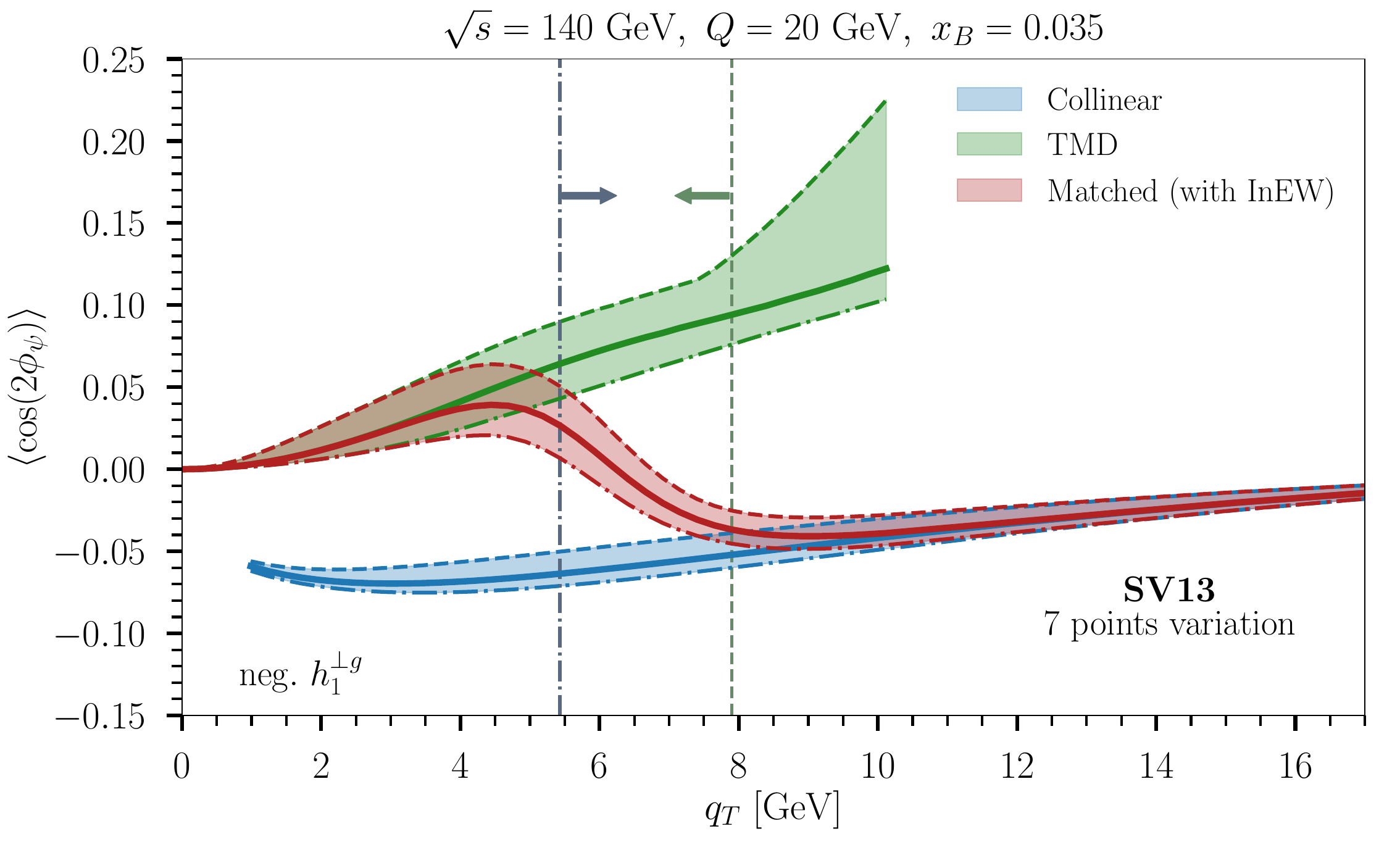}}
  \caption{\it Same as Fig.~\ref{fig: matched asymmetries (140) - pos} but for a negative sign $h_1^{\perp g}$.}
  \label{fig: matched asymmetries (140) - neg}
\end{figure}

Finally, we consider the same Sudakov factor used for the isotropic term also for the numerator of the asymmetry in the TMD factorization. While this is correct for the perturbative Sudakov, that should apply to the whole cross section~\cite{Catani:2017tuc, Catani:2017tuc}, the nonperturbative Sudakov associated with the numerator of the asymmetry might be different from that used for the denominator.
Nevertheless, in this paper we will assume the nonperturbative Sudakov in the numerator of the asymmetry to be the same as in the denominator. This can be improved upon in the future after error bands have been reduced by means of EIC data.

Adopting the same PDF and LDME sets as in the previous section, in Fig.~\ref{fig: matched asymmetries (140) - pos} (Fig.~\ref{fig: matched asymmetries (140) - neg}) we present $\langle \cos(2\phi_\psi) \rangle$, assuming a positive (negative) $h_1^{\perp g}$ distribution, at $\sqrt s = 140~{\rm GeV}$ and for three different values of $Q$: $3~{\rm GeV}$,  $14~{\rm GeV}$, and $20~{\rm GeV}$. For each $Q$ we also fix $\xB$ as indicated above each panel, whereas $z$ has been integrated in the region $0<z<1$.
Although one might expect a monotonic decrease of the asymmetry size with increasing $Q$, the actual dependence on $Q$ varies with the LDME set adopted. The most striking example is observed at $Q = 3~{\rm GeV}$, where the comparison between Fig.~\ref{subfig:a-asy140-pos} and Fig.~\ref{subfig:d-asy140-pos} (or Fig.~\ref{subfig:a-asy140-neg} and Fig.~\ref{subfig:d-asy140-neg}) shows that the SV13 set leading to significantly smaller predictions compared to the CM12 one. This difference reduces when going to higher $Q$, as shown by Figs.~\ref{subfig:b-asy140-pos} and~\ref{subfig:e-asy140-pos} (or Figs.~\ref{subfig:b-asy140-neg} and~\ref{subfig:e-asy140-neg}) and by Figs.~\ref{subfig:c-asy140-pos} and~\ref{subfig:f-asy140-pos} (Figs.~\ref{subfig:c-asy140-neg} and~\ref{subfig:f-asy140-neg}). This observation is in agreement with the findings of \cite{Bor:2022fga, Bor:2025ztq}.
Another observation concerns the choice of the $h_1^{\perp g}$ sign, which visibly affects the behavior of the matched curve. 
Explicitly, in Fig.~\ref{fig: matched asymmetries (140) - neg} the negative sign of $h_1^{\perp g}$ inevitably leads to the presence of a node in the matched curve, whereas in Fig.~\ref{fig: matched asymmetries (140) - pos} we have that the TMD and collinear predictions agree in sign, with no node found in the matched curve.
We also point out that, despite generally having significantly larger uncertainties, the TMD curves correctly display the expected behavior at small $q_\T$, namely $\langle \cos(2\phi_\psi) \rangle \sim q_\T^2$. 

\begin{figure}[t]
  \centering    
  \subfloat[\label{subfig:a-asy140-BCO}]{\includegraphics[width=.47\linewidth, keepaspectratio]{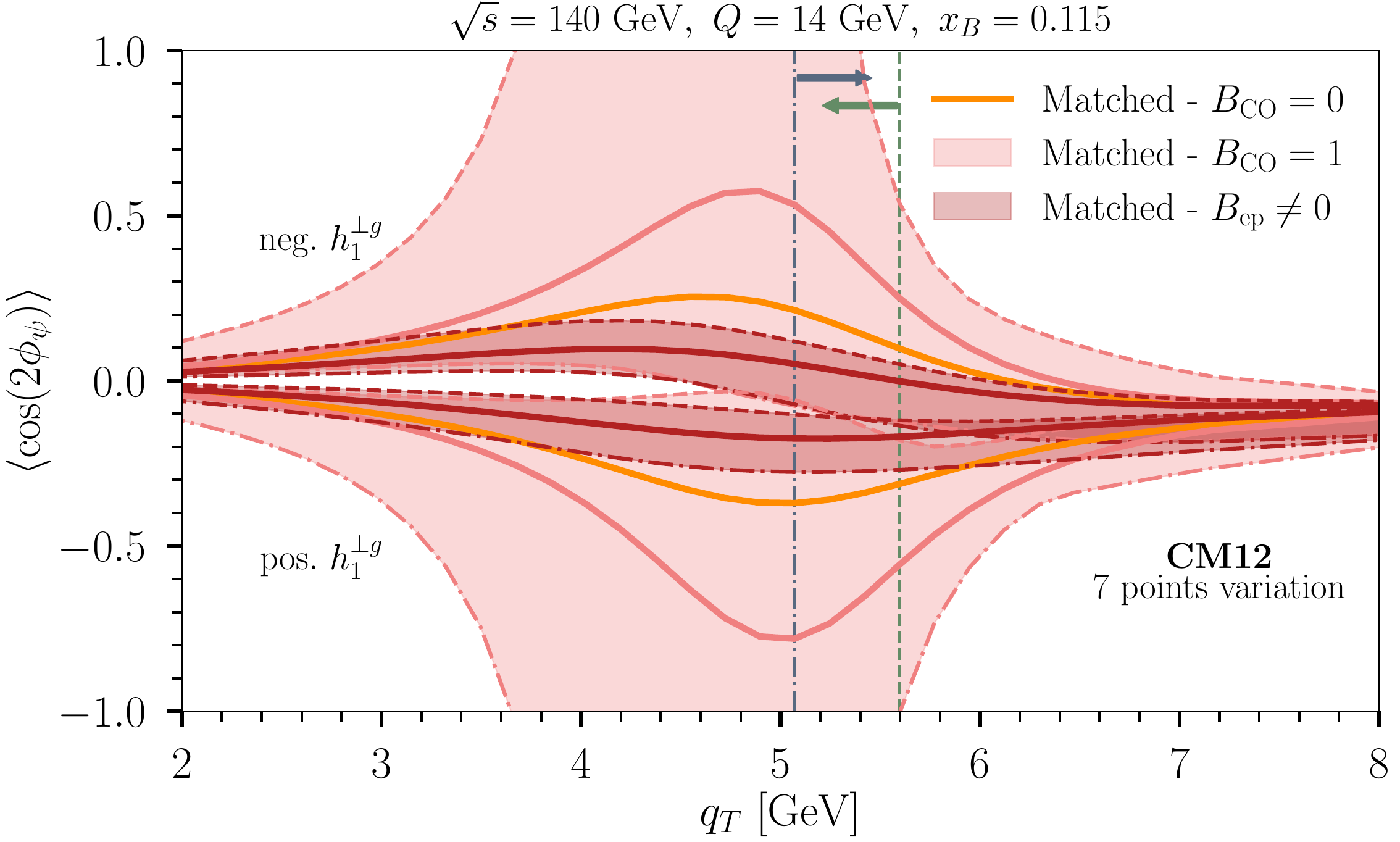}}
  \hfill
  \subfloat[\label{subfig:b-asy140-BCO}]{\includegraphics[width=.47\linewidth, keepaspectratio]{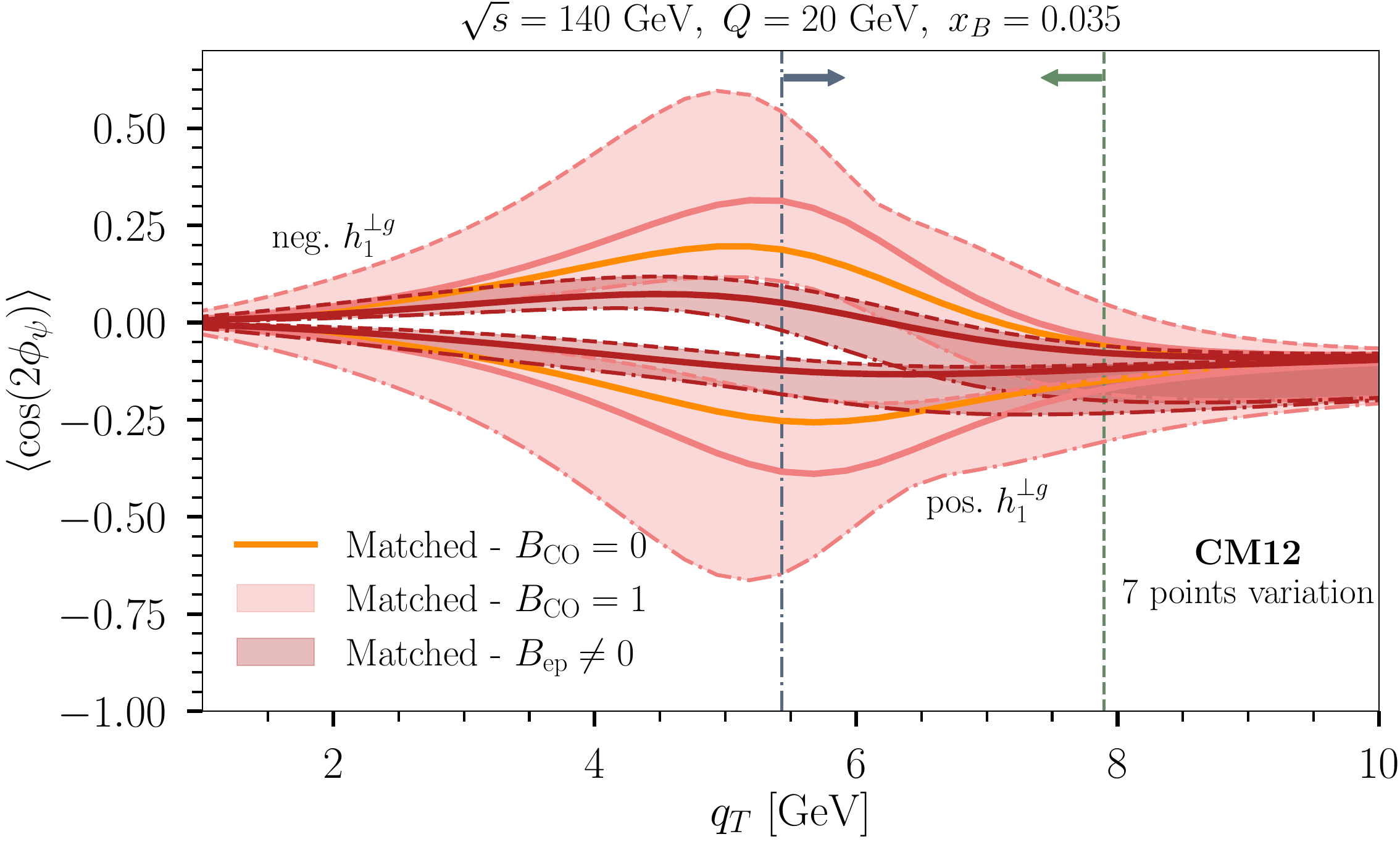}}\\
  \subfloat[\label{subfig:c-asy140-BCO}]{\includegraphics[width=.47\linewidth, keepaspectratio]{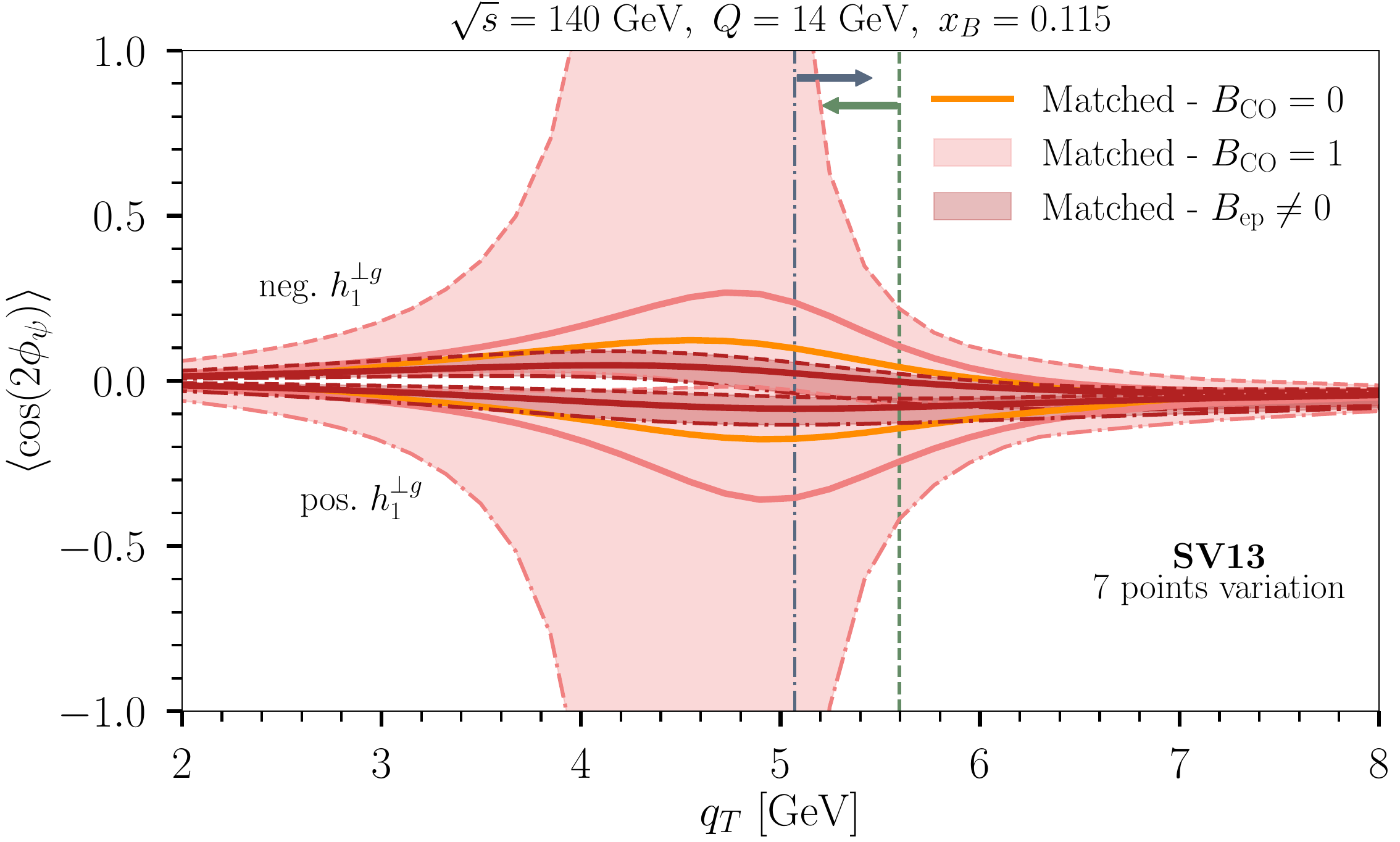}}
  \hfill
  \subfloat[\label{subfig:d-asy140-BCO}]{\includegraphics[width=.47\linewidth, keepaspectratio]{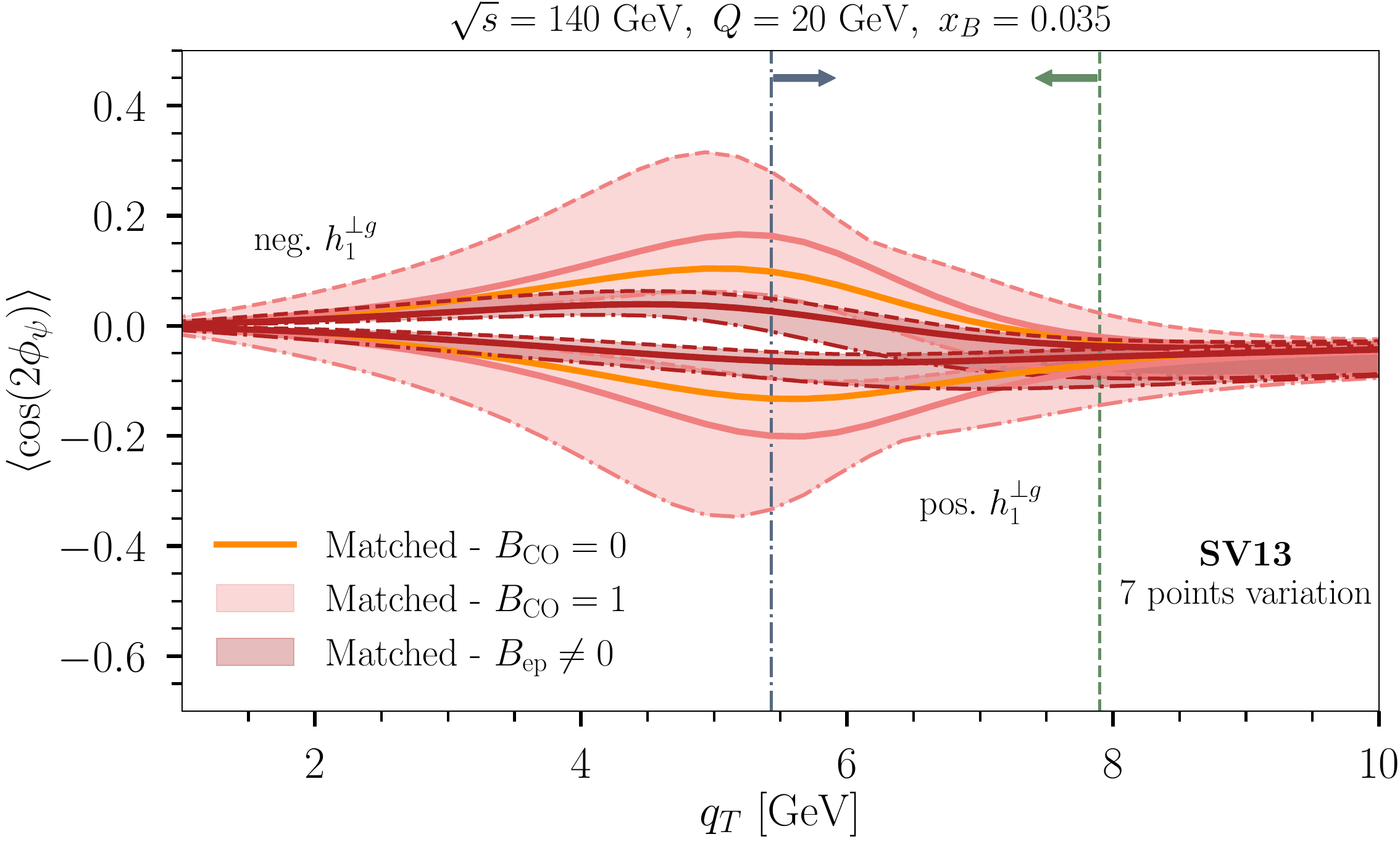}}
  \caption{\it Dependence of the $\langle \cos (2\phi_\psi) \rangle$ asymmetry on the $B_{\rm CO}$ choice, with $B_{\rm CO} = 1$ from Ref.~\cite{Echevarria:2024idp} and $B_{ep} \neq 0$ from Ref.~\cite{Boer:2023zit}. Asymmetries are shown with respect to $q_\T$ at fixed kinematics: (a) and (c) $Q = 14~{\rm GeV}$ and $\xB = 0.115$ ; (b) and (d) $Q = 20~{\rm GeV}$ and $\xB = 0.035$. Upper panels are evaluated using CM12, whereas SV12 is used for the lower ones. For clarity, we do not show the error band for the case $B_{\rm CO} = 0$, but only its central value, namely ${C_1 = C_2 = C_3 = 1}$. Positive and negative signs of the $h_1^{\perp g}$ distribution are shown together, as indicated by the labels.}
  \label{fig: TMD-ShF investigation - asymmetry}
\end{figure}

In Fig.~\ref{fig: TMD-ShF investigation - asymmetry} we investigate the effects of the TMD-ShF on the asymmetry.
We are presenting the two choices of the $h_1^{\perp g}$ distribution together, with the positive one producing the curves on the negative side, and vice versa.
As for the isotropic case, we compare the central value of the no-TMD-ShF case with the band obtained considering $B_{\rm CO} = 1$ (hence $B_{ep} = 0$) and $B_{ep} \neq 0$. To avoid the unphysical negative cross section observed for $B_{\rm CO} = 1$, we impose that the denominator of the asymmetry, when the TMD and FO contributions are combined, is always positive.
We focus on the higher energy only, $\sqrt s = 140~{\rm GeV}$, and we take both $Q = 14~{\rm GeV}$ and $Q = 20~{\rm GeV}$. 
Similarly to the isotropic case, it seems that $B_\psi$ has the effect to enhance the asymmetry compared to $B_{\rm CO} = 0$, whilst $B_{ep}$ has the opposite effect of suppressing the observable. The position of the peak of the distribution is instead comparable for all three cases. Although the bands obtained with the two TMD-ShFs are in better agreement with each other compared with the isotropic case, the $B_{\rm CO} = 1$ case presents an important shortcoming: the band of the matched curve overshoots $1$. The $q_\T$ range where this happens is comparable to when the cross section in the TMD framework becomes negative. Thus, we once again suspect that the extreme behavior shown by the $B_{\rm CO} = 1$ case suggests missing contributions.\footnote{While it may be possible to prevent asymmetries from overshooting unity by considering a different nonperturbative Sudakov factor for the numerator and denominator of the asymmetry, there still is the issue of the cross section becoming negative at some point within the TMD region. Since the two problems seem to occur in a similar $q_\T$ range, a more likely solution is that $B_{\rm CO} = 1$ (and to some extent $B_{\rm CO} = 0$ too) is simply not the correct choice.
However, a definitive exclusion of the $B_{\rm CO} = 1$ case requires a careful study of the subleading terms related to the approximation $z \approx 1$, as they might also affect the distribution at high $q_\T$.}

\section{Conclusions}
\label{sec: conclusions}

In this paper we have investigated the role of the TMD-ShF in the $q_\T$ spectra of the cross section and average $\cos (2\phi_\psi)$ asymmetry in $J/\psi$ production in SIDIS at the EIC. In particular, by means of the InEW matching method, we have obtained predictions for these observables valid for the full $q_\T$ range, such that there is compatibility with the TMD factorization description at low $q_\T$ and with collinear factorization at high $q_\T$. We have especially focused on the low and intermediate region, where the cross section is largest.
For the cross section we have considered two standard center of mass energies for the EIC, $\sqrt s = 45~{\rm GeV}$ and $\sqrt s = 140~{\rm GeV}$, whereas only the latter has been considered for the $\cos(2\phi_\psi)$ asymmetry because the former case is qualitatively very similar, although the overall magnitude of the observable drops faster with $Q$ at lower energy. This does not mean that such an observable cannot be measured at $\sqrt s = 45~{\rm GeV}$ as well; for instance, asymmetries up to $\sim10~\%$ are found for $Q = 12~{\rm GeV}$ and $\xB \sim 0.1$, with qualitatively the same shapes as at higher $\sqrt s$. In all cases, the results are obtained at fixed photon virtuality $Q$ and Bjorken $\xB$, while the $J/\psi$ fraction $z$ has been integrated over the region $0<z<1$. At low $q_\T$ the dominant contributions to both the cross section and asymmetry come from the high-$z$ region, as $z \to 1$ when $q_\T \to 0$. This allows us to draw conclusions about the novel TMD quantities that are considered for $z = 1$ only, despite the integration over $z$. In particular, we have observed significant differences in the predictions depending on the perturbative tail of the TMD-ShF employed, about which there has been debate in the literature.

For the cross section, we observe that the $B_{\rm \psi} = 1$ term causes an enhancement of the distribution at $q_\T \ll \mu$ compared with the case where there are no TMD effects associated with the outgoing particle ($B_{\rm CO} = 0$). On the other hand, $B_{ep}$, which explicitly depends on $Q$, has the opposite effect, suppressing the distribution in the same $q_\T$ region. These differences are expected to become larger with increasing $Q$. In particular, the prediction band that solely includes $B_{\psi}$ is found incompatible with the one that also includes $B_{ep}$ for sufficiently large $Q$. Hence, data taken at large $Q$ may distinguish the presence of nontrivial process dependence contributions (namely $B_{ep}$) in the perturbative Sudakov factor.
Of course, the data, and in particular the position of the peak of the distribution, is sensitive to the nonperturbative content in the Sudakov factor too. However, at present variations due to the $J/\psi$ contribution to the nonperturbative Sudakov factor are much smaller compared to the other sources of uncertainty considered here, namely LDME uncertainties, as well as scale and $A_{\rm NP}$ variations.

Regarding $\langle \cos (2\phi_\psi) \rangle$, its behavior depends on the sign of the linearly polarized TMD distribution of the proton. In particular, the presence (absence) of a node in the distribution at intermediate $q_\T$ is related to the positive (negative) sign of the distribution.
Moreover, as for the cross section, also $\langle \cos (2\phi_\psi) \rangle$ is sensitive to the TMD-ShF used, with $B_\psi$ leading to potentially greater asymmetries, whereas $B_{ep}$ leads to a suppression of them.
We also observe another major difference between the inclusion or exclusion of $B_{ep}$ from $B_{\rm CO}$, which is the edge of the matched curve band overshooting $1$. This finding, which is driven by the TMD contribution, may suggest the need for a more suppressing nonperturbative Sudakov factor, and/or that corrections to the approximation $z=1$ may be larger than anticipated, leading to a suppression of the asymmetry at $\Lambda_{\rm QCD} \ll q_\T \lesssim \mu/2$. However, we suspect that the solution to this unphysical behavior is simply that $B_{ep} \neq 0$, in support of the analytic calculations in \cite{Boer:2023zit}.

In conclusion, both cross section and $\cos(2\phi_\psi)$ asymmetry in $J/\psi$ production in SIDIS have valuable information regarding both the partonic (more specifically gluonic) content of protons, but also the production mechanism of the quarkonium. In particular, experimental data can test the validity of the factorization of such process and explore the presence of nontrivial process-dependent effects in the soft-gluon resummation. 

\section*{Acknowledgment}
We thank J.-P.~Lansberg for valuable discussions.
This work is supported by the ERC (grant 101041109 “BOSON”). Views and opinions expressed are however those of the authors only and do not necessarily reflect those of the European Union or the European Research Council Executive Agency. Neither the European Union nor the granting authority can be held responsible for them.
This project has also received funding from the European Union’s Horizon 2020 research and innovation programme under grant agreement No. 824093 (STRONG 2020) and is part of its JRA4-TMD-neXt Work-Package. This project has also received funding from the French Agence Nationale de la Recherche via the grant ANR-20-CE31-0015 (“PrecisOnium”) and was also partly supported by the French CNRS via the IN2P3 project GLUE@NLO.

\appendix
\section{The separation of single and double logarithms through auxiliary scales}
\label{app: anom dim}

In this appendix we revisit the separation of the single logarithms associated with the $J/\psi$ into a universal component and a process-dependent piece, as discussed in Ref.~\cite{Boer:2023zit}: ${\Delta_{ep} = \Delta_\psi \times S_{ep}}$, where $\Delta_{ep}$ denotes the full TMD-ShF obtained from matching in the $ep$ process, $\Delta_\psi$ its universal part and $S_{ep}$ the process-dependent soft factor.
Following the phenomenological approach to the TMD-ShF derivation, single- and double-logarithms associated to the soft gluon emission from both the incoming gluon and outgoing $J/\psi$ can be found in (Eq. (2.27) of \cite{Boer:2023zit})
\begin{equation}
    L(q_T) = C_A \left( \log\frac{M_\psi^2 + Q^2}{q_T^2} - 1 - \log\frac{M_\psi^2}{M_\psi^2 + Q^2} - \frac{\beta_0}{6} \right)\ .
\label{eq: single and double logs collinear}
\end{equation}
This expression poses two problems: 1) the $\log({M_\psi^2}/{M_\psi^2 + Q^2})$ inserts, a priori, a $Q^2$ dependence in the TMD-ShF; 2) the first term (that after Fourier transforming yields a double logarithm when multiplied by $1/q_\T^2$) looks like it is exclusively evaluated at a fixed scale, hence with the matching possible at $\mu^2 = M_\psi^2 + Q^2$ only.
Both problems can be addressed by separating the logarithms, which we illustrate here by considering two different auxiliary scales: $\mu_f$ and $\mu_\psi$:
\begin{equation}
\begin{aligned}
    \log\frac{M_\psi^2 + Q^2}{q_T^2} & \to \log\frac{\mu_f^2}{q_T^2} + \log\frac{M_\psi^2 + Q^2}{\mu_f^2}\ , \\
    \log\frac{M_\psi^2}{M_\psi^2 + Q^2} & \to \log\frac{M_\psi^2}{\mu_\psi^2} + \log\frac{\mu_\psi^2}{M_\psi^2 + Q^2}\ .
\end{aligned}
\end{equation}
By identifying $\mu_f$ as the renormalization scale $\mu$, $\log (\mu^2/q_\T^2)$ correctly corresponds to the gluon TMD-PDF double logarithm.
All the remaining terms are to single logarithms that should be connected to different TMD quantities, more specifically to their anomalous dimensions $\gamma_a$. By considering the aforementioned separation of $\Delta_{ep}$, we have three anomalous dimensions: $\gamma_f$ for the gluon TMD-PDF, $\gamma_\psi$ for $\Delta_\psi$ and $\gamma_S$ for $S_{ep}$.
The first one is known and can be obtained, for instance, from \cite{Echevarria:2015uaa}
\begin{equation}
    \gamma_f = C_A \left( \log\frac{\zeta_f}{\mu^2} - \frac{\beta_0}{6} \right)\ ,
\end{equation}
while $\gamma_\psi$ was instead recently obtained in \cite{Echevarria:2024idp}
\begin{equation}
    \gamma_\psi = C_A\, \left(\log \zeta_\psi - 1 - \log\frac{M_\psi^2}{\mu_\psi^2}\right)\ ,
\end{equation}
where an additional logarithm of quarkonium scales, $\log (M_\psi^2/\mu_\psi^2)$, has been included.
By selecting ${\zeta_f\, \zeta_\psi = M_\psi^2 + Q^2}$, one then obtains the unknown $\gamma_S$, given by
\begin{equation}
    \gamma_S = - C_A \log \frac{\mu_\psi^2}{M_\psi^2 + Q^2}\ .
\end{equation}
Since the scale $\mu_\psi$ is arbitrary (and actually unrelated to $\mu$), one can simply take $\mu_\psi = M_\psi$ to remove any factorization scale dependences from either $\gamma_\psi$ and $\gamma_S$, and thus from $B_{\rm CO}$. In contrast, in Ref.~\cite{Boer:2023zit} the final result displays a (spurious) dependence on $\mu$ due to a partial removal of the rapidity logarithm ($\log(\zeta_f \zeta_\psi/\mu^2)$).
Hence, the forms of $B_\psi$ and $B_{ep}$ presented in Eqs.~\eqref{eq: Bpsi definition} and~\eqref{eq: Bep definition}, respectively, are the preferred ones.

We also remark on the difference with the analysis in \cite{Echevarria:2024idp}, for which no anomalous dimensions besides $\gamma_f$ and $\gamma_\psi$ were observed, i.e.~$\gamma_S = 0$. The authors verified their finding with a cross-check based on the $J/\psi$ yield in photoproduction (see section 3.4 of \cite{Echevarria:2024idp}).
However, as $\gamma_S \to 0$ ($\mu_\psi \equiv M_\psi$) when $Q \to 0$, the combined expression of $\gamma_\psi$ and $\gamma_{ep}$ actually passes the same cross-check.
Thus, more conclusive tests must be considered to establish the presence or absence of $\gamma_S$, for instance, by studying $J/\psi$ production at small/intermediate $q_\T$ in SIDIS, as discussed in this paper.

\bibliography{References}
\bibliographystyle{apsrev4-1}

\end{document}